\documentclass[aps,prl,reprint,twocolumn,superscriptaddress,preprintnumbers,nofootinbib]{revtex4-1}

\usepackage{amsthm}
\usepackage{amsmath}
\usepackage{graphicx}
\usepackage{slashed}
\usepackage{amssymb}
\usepackage{float}
\usepackage[colorlinks=True, citecolor=blue, urlcolor=blue, linkcolor=blue]{hyperref}
\usepackage[braket,qm]{qcircuit}
\usepackage{lineno}

\newcommand*\diff{\mathop{}\!\mathrm{d}}

\newcommand{\be}{\begin{eqnarray}}
\newcommand{\ee}{\end{eqnarray}}
\newcommand{\ma}{\mathrm}

\newcommand{\bs}{\boldsymbol}
\newcommand{\Tr}{\mathrm{Tr}}

\newcommand{\Ncycle}{\ensuremath{N_{\rm{cycle}}}}

\begin{document}

\title{Quantum simulation of open quantum systems in heavy-ion collisions}

\author{Wibe A. de Jong}
\email{wadejong@lbl.gov}
\affiliation{Computational Research Division, Lawrence Berkeley National Laboratory, Berkeley, CA 94720, USA}

\author{Mekena Metcalf}
\email{mmetcalf@lbl.gov}
\affiliation{Computational Research Division, Lawrence Berkeley National Laboratory, Berkeley, CA 94720, USA}

\author{James Mulligan}
\email{james.mulligan@berkeley.edu}
\affiliation{Nuclear Science Division, Lawrence Berkeley National Laboratory, Berkeley, California 94720, USA}
\affiliation{Physics Department, University of California, Berkeley, CA 94720, USA}

\author{Mateusz P\l osko\'n}
\email{mploskon@lbl.gov}
\affiliation{Nuclear Science Division, Lawrence Berkeley National Laboratory, Berkeley, California 94720, USA}

\author{Felix Ringer}
\email{fmringer@lbl.gov}
\affiliation{Nuclear Science Division, Lawrence Berkeley National Laboratory, Berkeley, California 94720, USA}

\author{Xiaojun Yao}
\email{xjyao@mit.edu}
\affiliation{Center for Theoretical Physics, Massachusetts Institute of Technology, Cambridge, MA 02139, USA}

\date{\today}
\preprint{MIT-CTP/5247}
\begin{abstract}

We present a framework to simulate the dynamics
of hard probes such as heavy quarks or jets in a hot, strongly-coupled
quark-gluon plasma (QGP) on a quantum computer.
Hard probes in the QGP can be treated as open quantum systems governed in the Markovian limit by the Lindblad equation.
However, due to large computational costs, most current phenomenological calculations of hard probes evolving in
the QGP use semiclassical approximations of the quantum evolution. 
Quantum computation can mitigate these costs, and offers the potential for a fully quantum treatment with exponential speedup over classical techniques.
We report a simplified demonstration of our framework on IBM Q quantum devices,
and apply the Random Identity Insertion Method (RIIM) to account for \textsc{cnot} depolarization noise, in addition to measurement error mitigation.
Our work demonstrates the feasibility of simulating open quantum systems on current and near-term quantum devices,
which is of broad relevance to applications in nuclear physics, quantum information, and other fields.

\end{abstract}

\maketitle

{\it Introduction.} Considerable advancements in
quantum devices, such as qubit coherence times,
have recently been achieved~\cite{Devoret2013, annurev-conmatphys-031119-050605, doi:10.1063/1.5088164, google_supremacy}. 
Together with parallel progress in quantum algorithms and executable quantum software, nontrivial quantum computations can be carried out, including hybrid quantum-classical algorithms such as the variational quantum eigensolver~\cite{McClean_2016,Peruzzo_2014, Kandala2017, Rubin2020, PhysRevX.8.011021, Chong2017}
and fully quantum simulations of the unitary time evolution of closed quantum systems~\cite{PhysRevB.101.184305, Smith2019}.
In high energy and nuclear physics, a variety of quantum computing 
applications have emerged~\cite{Preskill_2018,Jordan:2011ne,Kaplan:2017ccd,Preskill:2018fag,Klco:2018zqz,Dumitrescu:2018njn,
Klco:2018kyo,Chang:2018uoc,Roggero:2019myu,Klco:2019evd,Roggero:2019srp,
Cloet:2019wre,Bauer:2019qxa,Mueller:2019qqj,Wei:2019rqy,Holland:2019zju,Avkhadiev:2019niu,
Shaw:2020udc,Liu:2020eoa,Kreshchuk:2020dla,Kharzeev:2020kgc,Klco:2020aud,DiMatteo:2020dhe,Davoudi:2020yln,Bepari:2020xqi}.
In particular, \emph{quantum simulation} can be applied to study dynamics of large size systems that are in principle intractable with classical methods.
To perform such simulations, quantum circuits compiled into single- and multi-qubit gates can be implemented on digital quantum computers.

Many physical systems of interest are not closed, but consist of
a subsystem interacting with an environment.
The dynamics of the subsystem can be formulated as an \emph{open quantum system}.
In the Markovian limit (in which the environment correlation time is much smaller than the subsystem relaxation time), the evolution of the subsystem is governed by
a generalization of the Schr\"odinger equation known as 
the Lindblad equation \cite{KOSSAKOWSKI1972247,Lindblad:1975ef,Gorini:1976cm}, where instead of keeping track of all of the environmental degrees of freedom, one only needs to record environment correlators that are relevant for the subsystem evolution.
 A key challenge in extending quantum simulation to open quantum systems
is that the Lindblad evolution is non-unitary. During the last decade, algorithms have been developed to overcome this issue, most of which couple the subsystem with auxiliary qubits (whose dimension can be significantly smaller than that of the environment) such that the whole system evolves unitarily~\cite{PhysRevA.83.062317,PhysRevA.91.062308,Wei:2016,cleve_et_al:LIPIcs:2017:7477,PhysRevLett.118.140403,PhysRevA.101.012328,PhysRevResearch.2.023214}.
More recently, simulations of open quantum systems have been carried out on real quantum devices, but without error mitigation~\cite{Hu:2019}.

In this letter, we focus on the application of quantum simulations of open quantum systems
to relativistic heavy-ion collisions (HICs). Experiments at the Relativistic Heavy Ion Collider (RHIC)
and the Large Hadron Collider (LHC) create a hot ($T\approx150-500$ MeV), short-lived ($t\approx 10\;\mathrm{fm}/c$) quark-gluon plasma (QGP)~\cite{PhysRevD.27.140, Arsene:2004fa,Adcox:2004mh,Back:2004je,Adams:2005dq,LHC1review,Braun-Munzinger:2015hba, TheBigPicture}.
The QGP is a deconfined phase of QCD matter believed to have existed shortly after the Big Bang~\cite{Weinberg:1977ji}. 
The properties of the QGP can be investigated using jets or heavy quarks~\cite{Adare:2010de,Sirunyan:2017isk,Adamczyk:2017yhe,Acharya:2019jyg,Aaboud:2018twu} that involve energy scales much larger than the QGP temperature (``hard probes").

The evolution of hard probes in the QGP
can be treated as an open system evolving in a hot medium. A fully field-theoretical description of hard probes in the medium is challenging and typically various approximations are made. Most studies employ semiclassical Boltzmann or Fokker-Planck (equivalent to Langevin) equations~\cite{Gossiaux:2008jv,Schenke:2009gb,Wang:2013cia,Cao:2016gvr,Cao:2015hia,Du:2017qkv,Ke:2018tsh,Yao:2020xzw}; 
semiclassical transport equations are leading order terms in the gradient expansion of the Wigner transformed Lindblad equation~\cite{Blaizot:2017ypk,Yao:2020eqy}.
Recently, several studies have applied Lindblad equations directly to investigate quarkonia~\cite{Young:2010jq,Akamatsu:2011se,Gossiaux:2016htk,Brambilla:2017zei,Yao:2018nmy,Miura:2019ssi,Sharma:2019xum,Brambilla:2020qwo} and jets~\cite{Vaidya:2020cyi,Vaidya:2020lih}, which are valid if the subsystem and environment are weakly coupled. It is expected that as the size of the subsystem increases (such as the jet radiation phase space, or the number of heavy quarks~\cite{Andronic:2007bi,Abada:2019lih} in the subsystem), solving Lindblad equations would challenge the limits of classical computation. Quantum computing offers a possibility to remove the constraint on the subsystem size, and go beyond the approximations made in semiclassical approaches. Moreover, quantum simulation may provide a solution to the notoriously difficult sign problem in classical lattice QCD calculations of real time observables~\cite{Jordan:2011ne,Jordan:2011ci,Martinez_2016,Haase:2020kaj} (the same problem can also appear in open QCD systems).

In this letter, we outline a formulation of the evolution of hard probes in the QGP as a Lindblad equation and explore how simulations on Noisy Intermediate Scale Quantum (NISQ~\cite{Preskill_2018}) devices can be used to advance theoretical studies of hard probes in the QGP. Using a quantum algorithm for simulating the Lindblad equation, we study a toy model on IBM~Q simulators and quantum devices, and implement error mitigation for measurement and two-qubit gate noise. We demonstrate that quantum algorithms simulating simple Lindblad evolution
are tractable on current and near-term devices, in terms of available
number of qubits, gate depth, and error rates. 

\begin{figure}[!t]
\includegraphics[scale=0.2]{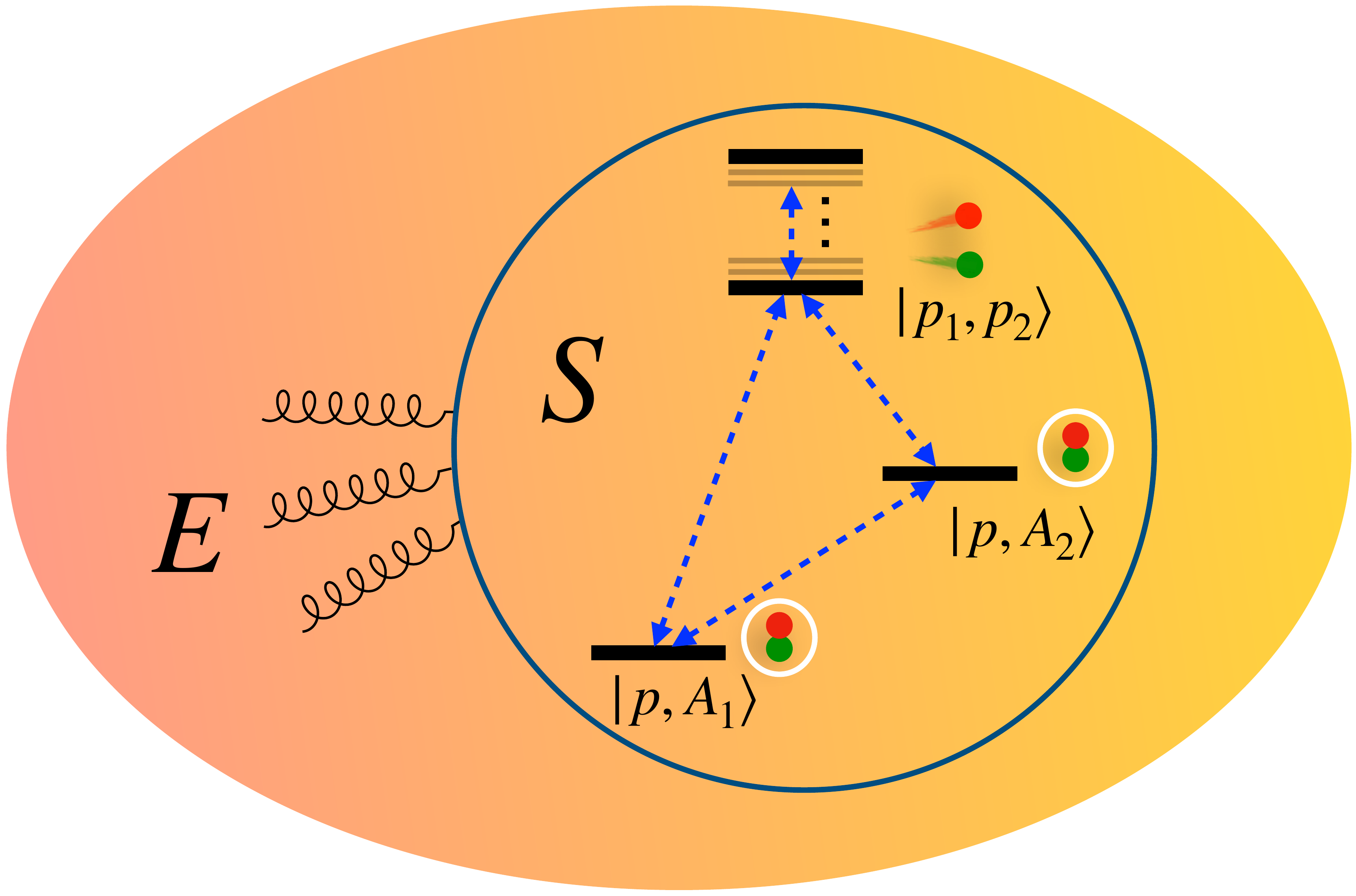}
\caption{A schematic illustration of a multi-level open quantum system $S$ interacting with a thermal environment $E$. The levels in $S$ can represent for example: (1) heavy quark-antiquark ($Q\bar{Q}$) bound states $|{\bs p},A_i\rangle$ with center-of-mass momentum ${\bs p}$ and quantum numbers $A_i$, and (2) unbound $Q\bar{Q}$ pairs $|{\bs p}_1, {\bs p}_2\rangle$ with momenta ${\bs p}_1, {\bs p}_2$. For jets the levels of $S$ can represent multi-parton states labeled by momenta $|p_1,\cdots,p_n \rangle$.~\label{fig:cartoon}}
\end{figure}

{\it Open quantum system formulation of hard probes in heavy-ion collisions.}
The Hamiltonian of the full system consisting of the hard probe (subsystem) and the QGP (environment) can be written as
\be
H &=& H_S + H_E + H_I\\
H_S &=& H_{S0} + H_{S1}\,.
\ee
Here $H_S$, $H_E$ and $H_I$ are the Hamiltonians of the subsystem, the
environment and their interaction, respectively. A schematic diagram of the setup is shown in Fig.~\ref{fig:cartoon}. We further split $H_S$ into
the free $H_{S0}$ and the interacting part of the subsystem $H_{S1}$. In quantum field theories, Hamiltonians are functionals of fields, which require discretization in position space~\cite{Preskill:2018fag}. Here, instead of simulating the dynamics of fields, we focus on simulating the dynamics of particle states, which is valid for hard probes. If we use multi-particle states $|p_1, A_1\rangle \otimes \cdots \otimes |p_n, A_n\rangle$ as the basis where $p_i$ is the four-momentum, $A_i$ represents all discrete quantum numbers, and $i=1,2,\ldots,n$, then both $H_{S0}$ and $H_{S1}$ are matrices and $H_{S0}$ is diagonal. Note that $H_{S1}$ is different from $H_I$: The former
is the interaction within the subsystem itself and independent of the
environment, while the latter represents the interaction between the subsystem and the
environment. For example, for jets in HICs, $H_{S1}$ can be
collinear radiation of collinear particles while $H_I$ can describe the Glauber
exchange between collinear particles (subsystem) and soft fields from the QGP
environment~\cite{Vaidya:2020cyi}.

The total density matrix of the subsystem and the
environment evolves under the von Neumann equation. In the interaction picture, this is given by
\be
\label{eqn:von_neumann}
\frac{\diff }{\diff t} \rho^{(\ma{int})}(t) = -i[H^{(\ma{int})}_I(t), \rho^{(\ma{int})}(t)]\,.
\ee
The operators are defined by
\be
\rho^{(\ma{int})}(t) &\equiv& e^{i (H_{S0} +H_E) t}   \rho(t)  e^{-i (H_{S0} +H_E) t} \\
H_{S1}^{(\ma{int})}(t) &\equiv& e^{i H_{S0} t} H_{S1} e^{-i H_{S0} t} \\
H_{I}^{(\ma{int})}(t) &\equiv& e^{i (H_{S0} +H_E) t}  H_I  e^{-i (H_{S0} +H_E) t}\,.
\ee
The interaction picture used here is special: it is the standard interaction picture for the subsystem but it is the Heisenberg picture for the environment. 
We will drop the superscript (int) from now on for simplicity but the reader should be reminded that we use the interaction picture throughout. We assume that the initial density matrix factorizes and the environment density matrix is a thermal state\footnote{The \emph{backreaction} of the QGP medium to jet energy
loss~\cite{CasalderreySolana:2004qm,Ruppert:2005uz,Chaudhuri:2005vc,Qin:2009uh,Betz:2010qh,Floerchinger:2014yqa,Chen:2017zte,Yan:2017rku,Tachibana:2020mtb,Casalderrey-Solana:2020rsj}, which may further
modify jet observables is beyond the scope of our considerations here. For a recent review, see Ref.~\cite{Cao:2020wlm}.}
\be
\rho(0) &=& \rho_S(0)\otimes \rho_E \\
\label{eqn:rho_E}
\rho_E &=& \frac{e^{-\beta H_E}}{\Tr(e^{-\beta H_E})}\,,
\ee
where $\beta=1/T$ is the inverse of the QGP temperature. 

After the
environment is traced out, the reduced evolution of the subsystem density
matrix is generally time-irreversible and non-unitary. If the coupling
between the subsystem and the environment is weak, the reduced evolution
equation can be cast as a Markovian Lindblad equation~\cite{KOSSAKOWSKI1972247,Lindblad:1975ef,Gorini:1976cm}:
\begin{align}\label{eqn:lindblad}
\frac{\diff}{\diff t} \rho_S(t)=
&
-i\big[H_{S1}(t) + H_L, \rho_S(t)\big] 
\nonumber\\&
+\sum_{j=1}^m \Big( L_j \rho_S(t) L_j^\dagger - \frac{1}{2}\big\{ L_j^\dagger L_j, \rho_S(t) \big\}  \Big) \,,
\end{align}
where $H_L$ denotes a thermal correction to $H_S$ generated by loop effects of $H_I$, and the $L_j$ are called Lindblad operators, whose explicit expressions
will be given for a toy model below. In general, if the dimension of the
subsystem is $d$, (i.e., $\rho_S(t)$ is a $d\times d$ matrix),
the number of independent Lindblad operators is $m = d^2-1$. When evaluating the Lindblad operators, an environment correlator of the form $\Tr_E[ O_E(t_1)O_E(t_2)\rho_E]$ is needed as input, where the $O_E$'s are some environment operators. This correlator can be evaluated perturbatively in thermal field theory if the environment is weakly-coupled. But the construction of the Lindblad equation only requires $H_I$ to be weak. In general $H_E$ itself can be strongly coupled, in which case the correlator has to be computed nonperturbatively using lattice QCD~\cite{Petreczky:2005nh,Banerjee:2011ra,Majumder:2012sh,Francis:2015daa,Brambilla:2020siz} or the AdS/CFT correspondence~\cite{Liu:2006ug,Liu:2006he,CasalderreySolana:2006rq,CaronHuot:2008uh,CasalderreySolana:2011us}. For the nonperturbative computation, one needs to formulate the theory such that the relevant correlator is gauge invariant, where effective field theory can be used. A concrete construction of gauge invariant correlators for quarkonium transport can be found in Refs.~\cite{Yao:2020eqy,Brambilla:2019tpt}.

\begin{figure}[!t]
\includegraphics[width = 0.48 \textwidth]{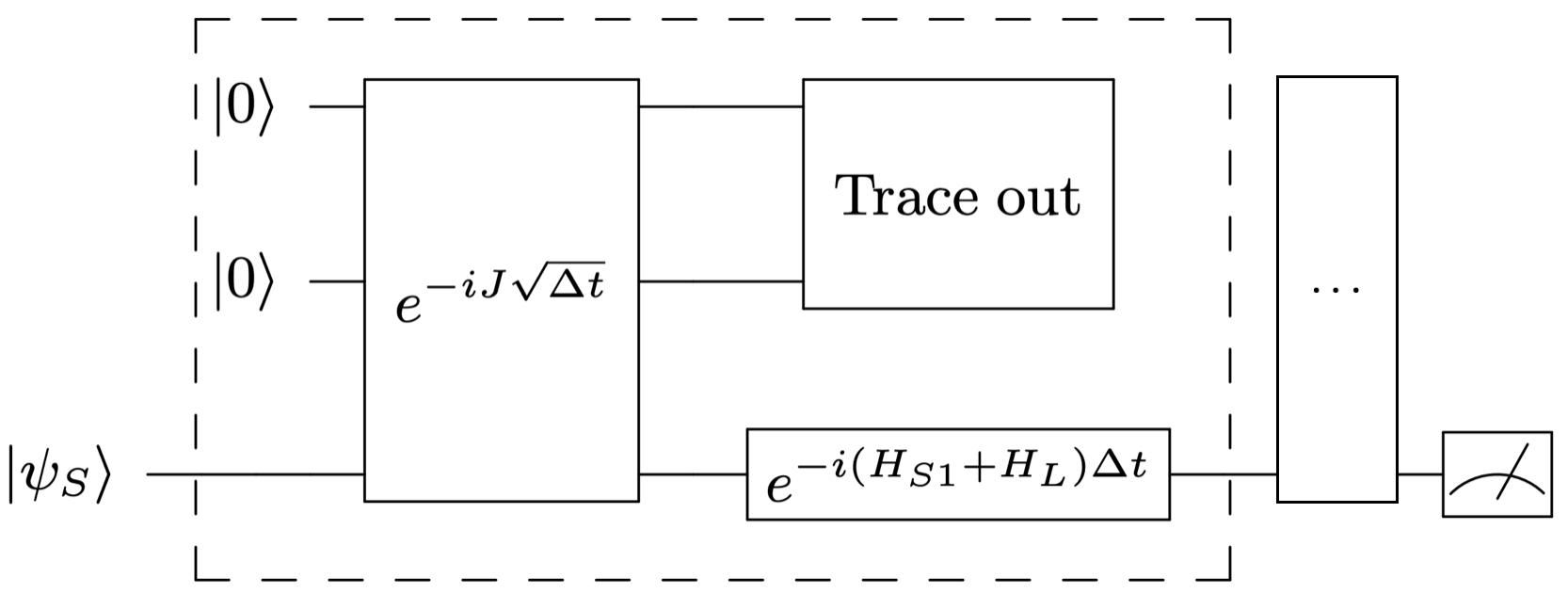}
\caption{Quantum algorithm to simulate Lindblad evolution based on the Stinespring dilation theorem.
The portion enclosed by the dashed line corresponds to one ``cycle'' 
of time $\Delta t$. Each cycle is repeated multiple times as indicated by the ellipsis in the box on the right. The measurement of the subsystem is performed at the end.~\label{fig:circuit}}
\end{figure}

{\it Quantum algorithm.}
We will apply a quantum algorithm based on the Stinespring dilation theorem, see for example Refs.~\cite{nielsen_chuang_2010,cleve_et_al:LIPIcs:2017:7477}, to simulate the Lindblad equation. The algorithm in terms of the evolution operators $J$, defined below, and $H_S$, is illustrated in Fig.~\ref{fig:circuit}.
The algorithm couples the subsystem with auxiliary qubits, which are traced out after each time step $\Delta t$.
The dimension of the auxiliary register is $m+1$ and the number of
qubits needed in practice for the register is $\ma{ceil}( \log_2(m+1)) \equiv \lceil{2\log_2d} \rceil$. Together with the number of qubits required to record the subsystem state, the total number of qubits needed is $\lceil{3\log_2d} \rceil$.
We use $\{ |0\rangle_a, |1\rangle_a \cdots, |m\rangle_a \}$ to label the basis of the auxiliary register, indicated by the subscript $a$. 

We assume the initial state $\rho_S(0) = |\psi_S(0)\rangle \langle \psi_S(0)|$ is a pure state\footnote{If it is a mixed state, then we decompose it into a linear
superposition of pure states. We just need to apply the circuit to each pure
state and take the linear superposition in the end.}.
At the beginning of each cycle at time $t$, the total density matrix of the subsystem and the auxiliary is set to be a $(m+1)\times(m+1)$ block matrix
\be
\rho(t) = |0\rangle_a \langle0|_a \otimes \rho_S(t) =\begin{pmatrix}
\rho_S(t) & 0 & \dots & 0\\
0 & 0 & \dots & 0\\
\vdots & \vdots &\ddots &\vdots\\
0 & 0 &\dots & 0
\end{pmatrix}\,.
\ee
The $J$-operator is also a $(m+1)\times(m+1)$ block matrix
\be
J = \begin{pmatrix}
0 & L_1^\dagger & \dots & L_m^\dagger\\
L_1 & 0 & \dots & 0\\
\vdots & \vdots &\ddots &\vdots\\
L_m & 0 &\dots & 0
\end{pmatrix}\,,
\ee
where each block is a $d \times d$ matrix. One can show that the circuit in Fig.~\ref{fig:circuit}
reproduces (\ref{eqn:lindblad}) when $\Delta t \to 0$. To simulate the evolution from $0$ to $t$, the size of the time steps is $\Delta t= t/N_{\ma{cycle}}$ where $N_{\ma{cycle}}$ is the number of cycles, see Fig.~\ref{fig:circuit}.

{\it Toy model and simulation on IBM~Q.}
Simulating real jets and heavy quarks on quantum devices requires a large number of fault-tolerant qubits. As a proof of concept, we consider the following toy model that includes qualitative features of hard probes:
\begin{align}
H_S &=\, H_{S0} = -\frac{\Delta E}{2}Z\\
H_E &=\, \int \diff^3 x \bigg[\frac{1}{2}\Pi^2 + \frac{1}{2}(\nabla \phi)^2 + \frac{1}{2}m^2\phi^2 + \frac{1}{4!}\lambda \phi^4 \bigg] \\
H_I &=\, gX\otimes \phi(x=0) \,,
\end{align}
where we use $X,Z$ to denote the single qubit Pauli gates (Pauli matrices). The subsystem Hamiltonian $H_S$ is a two level system with energy difference $\Delta E$. The two levels can correspond to the bound and unbound state of a heavy quark-antiquark pair, exchanging energy with QGP. The environment $H_E$ is a $3+1D$ scalar field theory, that together with (\ref{eqn:rho_E}) mimics the thermal QGP. Here $\Pi$ is the canonical momentum conjugate to $\phi$. The extension to gauge theories requires a gauge invariant formulation of the environment correlator as mentioned earlier. The environment correlator can be calculated nonperturbatively to all orders in $\lambda$. Here for simplicity, we set $m=\lambda =0$. Nonvanishing $m$ and $\lambda$ lead to different coefficients of the Lindblad operators but do not alter the quantum algorithm. The interaction strength $g$ between the subsystem and the environment is unitless.
In the Markovian limit, two Lindblad operators $j=0,1$ are relevant:
\begin{equation}
    L_{j}=\frac{\sqrt{\Gamma_j}}{2} (X - (-1)^j iY)\,,
\end{equation}
where $\Gamma_0=g^2 \Delta E n_B(\Delta E)/(2\pi)$, $\Gamma_1=g^2 \Delta E /(2\pi)+\Gamma_0$ and $n_B(\Delta E)=1/(\exp(\beta \Delta E)-1)$ is the Bose-Einstein distribution. We will neglect $H_L$ in this letter. For our numerical studies, we use a unit system where all quantities are counted in units of $T$, the temperature of the medium. We initialize the state as $\rho_S(t=0) = |0\rangle \langle0|$ and choose $\Delta E=1(T)$.

\begin{figure}[!t]
\includegraphics[scale=0.53]{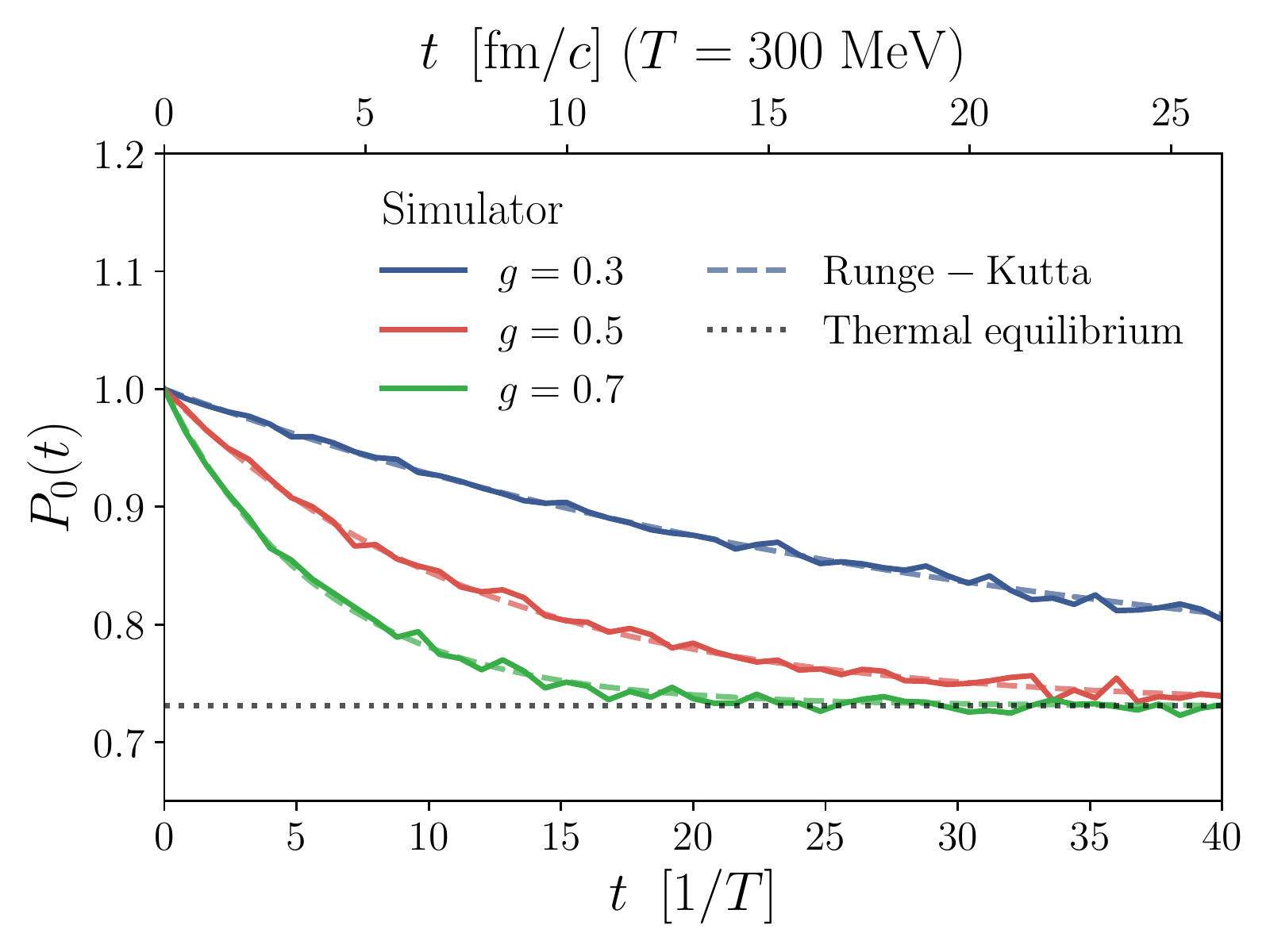}
\caption{Simulation of the quantum circuit with $\Ncycle=100$ for various system-environment couplings, along with numerical solution using a 4th order Runge-Kutta method. The upper time axis corresponds to a medium with a temperature of $T=300$~MeV. Each time point in the simulator result
consists of 80192 shots (runs).}
\label{fig:simulation_g}
\end{figure}

The result for this toy model obtained from the IBM~Q \texttt{qiskit} simulator~\cite{Qiskit} is shown in Fig.~\ref{fig:simulation_g}. We measure $P_0(t)\equiv \langle 0 | \rho_S(t) | 0\rangle $, which can be interpreted as the time-dependent nuclear modification factor.
Each time point corresponds to an independent quantum circuit, where the measurement
is performed only at the end, as shown in Fig. \ref{fig:circuit}.
The results of the quantum algorithm with
$\Ncycle{}=100$ are shown for different values of the coupling $g$. They are 
consistent with the results obtained with a 4th order Runge-Kutta method that solves Eq.~(\ref{eqn:lindblad}) classically. This agreement demonstrates that the circuit successfully 
solves the Lindblad equation. As expected, the strength of the coupling $g$ controls the rate of approaching thermalization.

In order to run the circuit on a quantum device, we select $\Ncycle{}=1$ in order to achieve a sufficiently small circuit depth. Modern quantum software packages are available to compile quantum circuits that approximate general unitary operators with minimal error and optimal depth ~\cite{Chong2017,younis2020qfast,davis2019heuristics, qsearch_placeholder}.
We synthesize a circuit for the $e^{-iJ\sqrt{\Delta t}}$ operator in terms of single qubit and \textsc{cnot} gates using the \texttt{qsearch} compiler~\cite{qsearch_placeholder}. The compiler yields circuits with 70 gates on average, including approximately 10 \textsc{cnot}s per cycle; an example circuit for one cycle is shown in the supplemental material.

\begin{figure}[!t]
\includegraphics[scale=0.53]{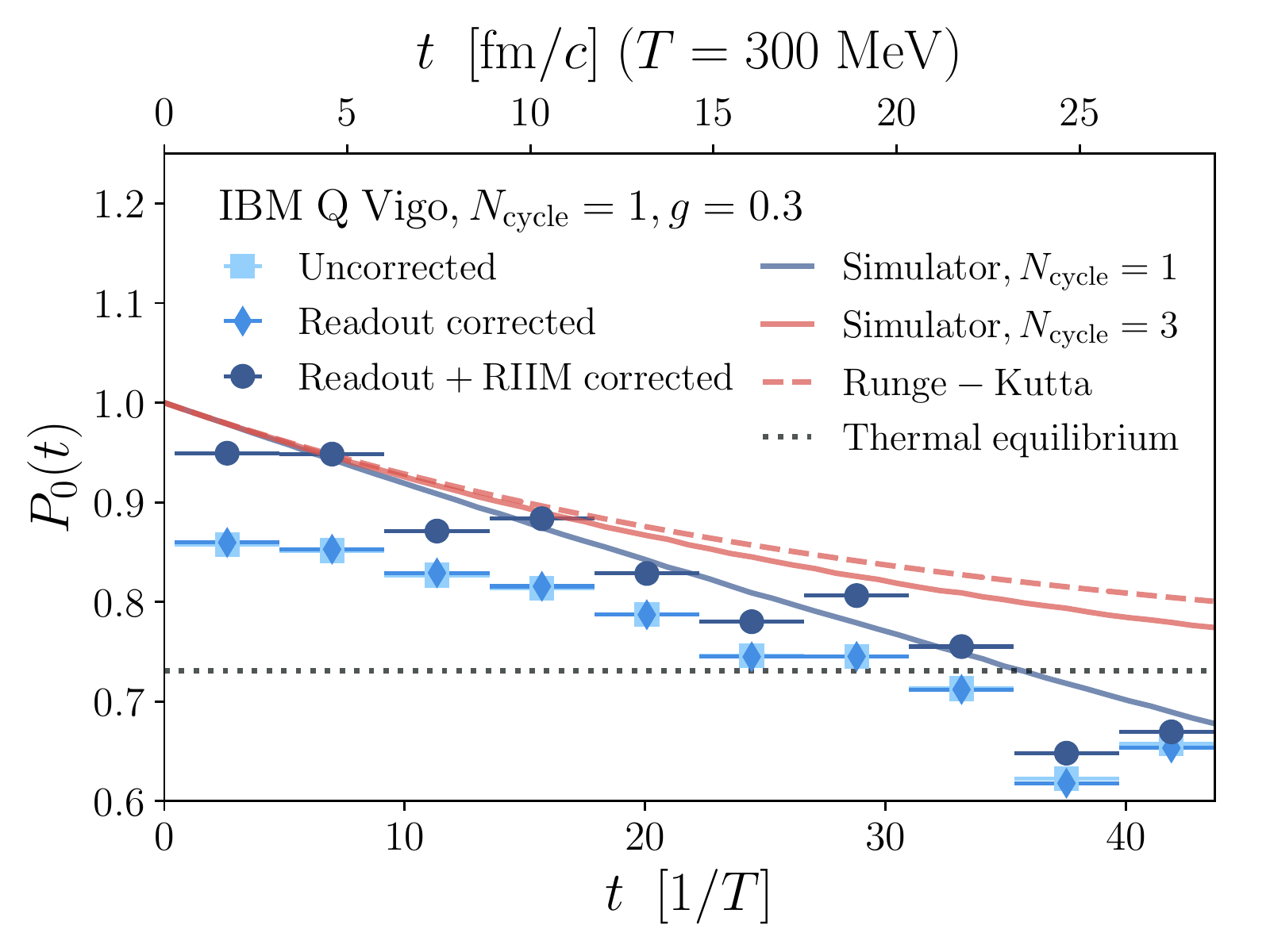}
\caption{Results from the IBM~Q Vigo device including different error mitigations compared to results from the \texttt{qiskit} simulator for $N_{\rm cycle}=1$ and $N_{\rm cycle}=3$ and the Runge-Kutta method. Higher values of $N_{\rm cycle}$ quickly converge to the result using the Runge-Kutta method.
Each time point in the simulator result
consists of 800192 shots (runs).}
\label{fig:device_result}
\end{figure}

The results obtained from IBM~Q Vigo device~\cite{IBMQVigo} are shown in Fig.~\ref{fig:device_result}.
In addition to the uncorrected result, the results with readout and \textsc{cnot} error mitigation are also shown. 
We correct 
the readout error using the constrained matrix inversion approach in IBM's \texttt{qiskit-ignis} package. The response matrix can be found in the supplemental material. 
We also correct for \textsc{cnot} noise using a leading order zero-noise extrapolation based on the recently developed resource efficient Random Identity Insertion Method (RIIM)~\cite{He:2020udd}. This procedure corrects for depolarization noise using a set of additional (\textsc{cnot})$^2$ identity insertions, at the expense of amplifying statistical noise.
Each data point corresponds to 5 evenly spaced time points that are averaged together.
Each time point is calculated from the average of 49152 shots (runs). We observe that the error mitigation is more important at small values of $t$.
Similar results were reproduced on the IBM~Q Valencia and Santiago devices~\cite{IBMQValencia, IBMQSantiago}.

Overall, we observe good agreement of the results from the quantum device with the results from the simulator for $\Ncycle{}=1$ after the error mitigation is applied.
The choice of $\Ncycle{}=1$ is seen to be
a reasonable approximation for sufficiently small $t$. 
Moreover, a modest increase to 
$\Ncycle{}=3$, as shown by the simulator in Fig.~\ref{fig:device_result}, yields 
considerably improved convergence, which is promising for near-term applications.
These results demonstrate that the simulation of open quantum system dynamics relevant for HICs should be feasible on current and near-term quantum devices.

{\it Conclusions and Outlook.} We performed simulations of open quantum systems using quantum devices from IBM~Q. In particular, we focused on simulating the non-unitary evolution of a subsystem governed by the Lindblad equation. We demonstrated that digital quantum simulations with a few qubits and a circuit depth of $\sim 70$ gate operations with $\sim 10$ \textsc{cnot} gates are feasible on current quantum devices. We used the \texttt{qsearch} compiler to construct the quantum circuit, and implemented two-qubit gate error mitigation using zero noise extrapolation with the Random Identity Insertion Method (RIIM), in addition to readout error mitigation. 
Simulating open quantum systems is of great importance for theoretical studies of hard probes in heavy-ion collisions.
The open quantum system formulation allows one to go beyond semiclassical transport calculations currently used in most phenomenological studies. Future calculations, using a time dependent environment density matrix may allow one to explore a broad range of physical models by varying medium properties such as the initial temperature, microscopic structure, or the probe-medium coupling. 
Open quantum systems are also relevant for various other systems in nuclear and high-energy physics such as studies of Cold Nuclear Matter effects at the future Electron-Ion Collider~\cite{Accardi:2012qut}, the resummation of large logarithms relevant for jet physics~\cite{Dasgupta:2001sh,Banfi:2002hw,Nagy:2007ty,Neill:2015nya} and studies of the Color Glass Condensate~\cite{Armesto:2019mna,Li:2020bys}.

\begin{acknowledgements}
{\it Acknowledgements.} We thank Christian Bauer, Volker Koch, Ben Nachman, Long-Gang Pang, Krishna Rajagopal, Phiala Shanahan, Rishi Sharma, Ramona Vogt, Xin-Nian Wang and Feng Yuan for helpful discussions. We acknowledge use of IBM Q for this work. The views expressed are those of the authors and do not reflect the official policy or position of IBM or the IBM Q team. 
JM, MP are supported by the U.S. Department of Energy, Office of Science, 
Office of Nuclear Physics, under the contract DE-AC02-05CH11231.
FR is supported by LDRD funding from Berkeley Lab provided by the U.S. Department of Energy under Contract No. DE-AC02-05CH11231. XY is supported by the U.S. Department of Energy, Office of Science, Office of Nuclear Physics under grant Contract Number DE-SC0011090. MM and WDJ were supported by the U.S. Department of Energy, Office of Science, Office of Advanced Scientific Computing Research Accelerated Research in Quantum Computing program under contract DE-AC02-05CH11231. This research used resources of the Oak Ridge Leadership Computing Facility, which is a User Facility supported  by U.S. Department of Energy, Office of Science, under Contract No. DE-AC05-00OR22725.
\end{acknowledgements}

\bibliographystyle{utphys}
\bibliography{main.bib}

\providecommand{\href}[2]{#2}\begingroup\raggedright\begin{thebibliography}{100}

\bibitem{Devoret2013}
M.~Devoret and R.~Schoelkopf, ``Superconducting circuits for quantum
  information: An outlook,''
  \href{http://dx.doi.org/10.1126/science.1231930}{{\em Science (New York,
  N.Y.)} {\bfseries 339} (03, 2013) 1169--74}.

\bibitem{annurev-conmatphys-031119-050605}
M.~Kjaergaard, M.~E. Schwartz, J.~Braumüller, P.~Krantz, J.~I.-J. Wang,
  S.~Gustavsson, and W.~D. Oliver, ``Superconducting qubits: Current state of
  play,''
  \href{http://dx.doi.org/10.1146/annurev-conmatphys-031119-050605}{{\em Annual
  Review of Condensed Matter Physics} {\bfseries 11} no.~1, (2020) 369--395}.

\bibitem{doi:10.1063/1.5088164}
C.~D. Bruzewicz, J.~Chiaverini, R.~McConnell, and J.~M. Sage, ``Trapped-ion
  quantum computing: Progress and challenges,''
  \href{http://dx.doi.org/10.1063/1.5088164}{{\em Applied Physics Reviews}
  {\bfseries 6} no.~2, (2019) 021314}.

\bibitem{google_supremacy}
F.~Arute, K.~Arya, R.~Babbush, D.~Bacon, J.~C. Bardin, R.~Barends, R.~Biswas,
  S.~Boixo, F.~G. S.~L. Brandao, and D.~A. e.~a. Buell, ``Quantum supremacy
  using a programmable superconducting processor,''
  \href{http://dx.doi.org/10.1038/s41586-019-1666-5}{{\em Nature} {\bfseries
  574} no.~7779, (2019) 505--510}.
  \url{https://doi.org/10.1038/s41586-019-1666-5}.

\bibitem{McClean_2016}
J.~R. McClean, J.~Romero, R.~Babbush, and A.~Aspuru-Guzik, ``The theory of
  variational hybrid quantum-classical algorithms,''
  \href{http://dx.doi.org/10.1088/1367-2630/18/2/023023}{{\em New Journal of
  Physics} {\bfseries 18} no.~2, (Feb, 2016) 023023}.

\bibitem{Peruzzo_2014}
A.~Peruzzo, J.~McClean, P.~Shadbolt, M.-H. Yung, X.-Q. Zhou, P.~J. Love,
  A.~Aspuru-Guzik, and J.~L. O’Brien, ``A variational eigenvalue solver on a
  photonic quantum processor,''
  \href{http://dx.doi.org/10.1038/ncomms5213}{{\em Nature Communications}
  {\bfseries 5} no.~1, (Jul, 2014) }.

\bibitem{Kandala2017}
A.~Kandala, A.~Mezzacapo, K.~Temme, M.~Takita, M.~Brink, J.~M. Chow, and J.~M.
  Gambetta, ``Hardware-efficient variational quantum eigensolver for small
  molecules and quantum magnets,''
  \href{http://dx.doi.org/10.1038/nature23879}{{\em Nature} {\bfseries 549}
  no.~7671, (2017) 242--246}. \url{https://doi.org/10.1038/nature23879}.

\bibitem{Rubin2020}
{Google AI Quantum and collaborators}, ``Hartree-fock on a superconducting
  qubit quantum computer,''
  \href{http://dx.doi.org/10.1126/science.abb9811}{{\em Science} {\bfseries
  369} (08, 2020) 1084--1089}.

\bibitem{PhysRevX.8.011021}
J.~I. Colless, V.~V. Ramasesh, D.~Dahlen, M.~S. Blok, M.~E. Kimchi-Schwartz,
  J.~R. McClean, J.~Carter, W.~A. de~Jong, and I.~Siddiqi, ``Computation of
  molecular spectra on a quantum processor with an error-resilient algorithm,''
  \href{http://dx.doi.org/10.1103/PhysRevX.8.011021}{{\em Phys. Rev. X}
  {\bfseries 8} (Feb, 2018) 011021}.
  \url{https://link.aps.org/doi/10.1103/PhysRevX.8.011021}.

\bibitem{Chong2017}
F.~T. Chong, D.~Franklin, and M.~Martonosi, ``Programming languages and
  compiler design for realistic quantum hardware,''
  \href{http://dx.doi.org/10.1038/nature23459}{{\em Nature} {\bfseries 549}
  (09, 2017) 180--187}.

\bibitem{PhysRevB.101.184305}
L.~Bassman, K.~Liu, A.~Krishnamoorthy, T.~Linker, Y.~Geng, D.~Shebib,
  S.~Fukushima, F.~Shimojo, R.~K. Kalia, A.~Nakano, and P.~Vashishta, ``Towards
  simulation of the dynamics of materials on quantum computers,''
  \href{http://dx.doi.org/10.1103/PhysRevB.101.184305}{{\em Phys. Rev. B}
  {\bfseries 101} (May, 2020) 184305}.
  \url{https://link.aps.org/doi/10.1103/PhysRevB.101.184305}.

\bibitem{Smith2019}
A.~Smith, M.~S. Kim, F.~Pollmann, and J.~Knolle, ``Simulating quantum many-body
  dynamics on a current digital quantum computer,''
  \href{http://dx.doi.org/10.1038/s41534-019-0217-0}{{\em npj Quantum
  Information} {\bfseries 5} (11, 2019) 106}.

\bibitem{Preskill_2018}
J.~Preskill, ``Quantum computing in the nisq era and beyond,''
  \href{http://dx.doi.org/10.22331/q-2018-08-06-79}{{\em Quantum} {\bfseries 2}
  (Aug, 2018) 79}. \url{http://dx.doi.org/10.22331/q-2018-08-06-79}.

\bibitem{Jordan:2011ne}
S.~P. Jordan, K.~S. Lee, and J.~Preskill, ``{Quantum Algorithms for Quantum
  Field Theories},'' \href{http://dx.doi.org/10.1126/science.1217069}{{\em
  Science} {\bfseries 336} (2012) 1130--1133},
  \href{http://arxiv.org/abs/1111.3633}{{\ttfamily arXiv:1111.3633
  [quant-ph]}}.

\bibitem{Kaplan:2017ccd}
D.~B. Kaplan, N.~Klco, and A.~Roggero, ``{Ground States via Spectral Combing on
  a Quantum Computer},'' \href{http://arxiv.org/abs/1709.08250}{{\ttfamily
  arXiv:1709.08250 [quant-ph]}}.

\bibitem{Preskill:2018fag}
J.~Preskill, ``{Simulating quantum field theory with a quantum computer},''
  \href{http://dx.doi.org/10.22323/1.334.0024}{{\em PoS} {\bfseries
  LATTICE2018} (2018) 024}, \href{http://arxiv.org/abs/1811.10085}{{\ttfamily
  arXiv:1811.10085 [hep-lat]}}.

\bibitem{Klco:2018zqz}
N.~Klco and M.~J. Savage, ``{Digitization of scalar fields for quantum
  computing},'' \href{http://dx.doi.org/10.1103/PhysRevA.99.052335}{{\em Phys.
  Rev. A} {\bfseries 99} no.~5, (2019) 052335},
  \href{http://arxiv.org/abs/1808.10378}{{\ttfamily arXiv:1808.10378
  [quant-ph]}}.

\bibitem{Dumitrescu:2018njn}
E.~Dumitrescu, A.~McCaskey, G.~Hagen, G.~Jansen, T.~Morris, T.~Papenbrock,
  R.~Pooser, D.~Dean, and P.~Lougovski, ``{Cloud Quantum Computing of an Atomic
  Nucleus},'' \href{http://dx.doi.org/10.1103/PhysRevLett.120.210501}{{\em
  Phys. Rev. Lett.} {\bfseries 120} no.~21, (2018) 210501},
  \href{http://arxiv.org/abs/1801.03897}{{\ttfamily arXiv:1801.03897
  [quant-ph]}}.

\bibitem{Klco:2018kyo}
N.~Klco, E.~Dumitrescu, A.~McCaskey, T.~Morris, R.~Pooser, M.~Sanz, E.~Solano,
  P.~Lougovski, and M.~Savage, ``{Quantum-classical computation of Schwinger
  model dynamics using quantum computers},''
  \href{http://dx.doi.org/10.1103/PhysRevA.98.032331}{{\em Phys. Rev. A}
  {\bfseries 98} no.~3, (2018) 032331},
  \href{http://arxiv.org/abs/1803.03326}{{\ttfamily arXiv:1803.03326
  [quant-ph]}}.

\bibitem{Chang:2018uoc}
C.~C. Chang, A.~Gambhir, T.~S. Humble, and S.~Sota, ``{Quantum annealing for
  systems of polynomial equations},''
  \href{http://dx.doi.org/10.1038/s41598-019-46729-0}{{\em Sci. Rep.}
  {\bfseries 9} no.~1, (2019) 10258},
  \href{http://arxiv.org/abs/1812.06917}{{\ttfamily arXiv:1812.06917
  [quant-ph]}}.

\bibitem{Roggero:2019myu}
A.~Roggero, A.~C. Li, J.~Carlson, R.~Gupta, and G.~N. Perdue, ``{Quantum
  Computing for Neutrino-Nucleus Scattering},''
  \href{http://dx.doi.org/10.1103/PhysRevD.101.074038}{{\em Phys. Rev. D}
  {\bfseries 101} no.~7, (2020) 074038},
  \href{http://arxiv.org/abs/1911.06368}{{\ttfamily arXiv:1911.06368
  [quant-ph]}}.

\bibitem{Klco:2019evd}
N.~Klco, J.~R. Stryker, and M.~J. Savage, ``{SU(2) non-Abelian gauge field
  theory in one dimension on digital quantum computers},''
  \href{http://dx.doi.org/10.1103/PhysRevD.101.074512}{{\em Phys. Rev. D}
  {\bfseries 101} no.~7, (2020) 074512},
  \href{http://arxiv.org/abs/1908.06935}{{\ttfamily arXiv:1908.06935
  [quant-ph]}}.

\bibitem{Roggero:2019srp}
A.~Roggero and A.~Baroni, ``{Short-depth circuits for efficient expectation
  value estimation},''
  \href{http://dx.doi.org/10.1103/PhysRevA.101.022328}{{\em Phys. Rev. A}
  {\bfseries 101} no.~2, (2020) 022328},
  \href{http://arxiv.org/abs/1905.08383}{{\ttfamily arXiv:1905.08383
  [quant-ph]}}.

\bibitem{Cloet:2019wre}
I.~C. Cloët {\em et~al.}, ``{Opportunities for Nuclear Physics \& Quantum
  Information Science},'' in {\em {Intersections between Nuclear Physics and
  Quantum Information}}, I.~C. Cloët and M.~R. Dietrich, eds.
\newblock 3, 2019.
\newblock \href{http://arxiv.org/abs/1903.05453}{{\ttfamily arXiv:1903.05453
  [nucl-th]}}.

\bibitem{Bauer:2019qxa}
C.~W. Bauer, W.~A. de~Jong, B.~Nachman, and D.~Provasoli, ``{Quantum Algorithm
  for High Energy Physics Simulations},''
  \href{http://dx.doi.org/10.1103/PhysRevLett.126.062001}{{\em Phys. Rev.
  Lett.} {\bfseries 126} no.~6, (2021) 062001},
  \href{http://arxiv.org/abs/1904.03196}{{\ttfamily arXiv:1904.03196
  [hep-ph]}}.

\bibitem{Mueller:2019qqj}
N.~Mueller, A.~Tarasov, and R.~Venugopalan, ``{Deeply inelastic scattering
  structure functions on a hybrid quantum computer},''
  \href{http://dx.doi.org/10.1103/PhysRevD.102.016007}{{\em Phys. Rev. D}
  {\bfseries 102} no.~1, (2020) 016007},
  \href{http://arxiv.org/abs/1908.07051}{{\ttfamily arXiv:1908.07051
  [hep-th]}}.

\bibitem{Wei:2019rqy}
A.~Y. Wei, P.~Naik, A.~W. Harrow, and J.~Thaler, ``{Quantum Algorithms for Jet
  Clustering},'' \href{http://dx.doi.org/10.1103/PhysRevD.101.094015}{{\em
  Phys. Rev. D} {\bfseries 101} no.~9, (2020) 094015},
  \href{http://arxiv.org/abs/1908.08949}{{\ttfamily arXiv:1908.08949
  [hep-ph]}}.

\bibitem{Holland:2019zju}
E.~T. Holland, K.~A. Wendt, K.~Kravvaris, X.~Wu, W.~Erich~Ormand, J.~L. DuBois,
  S.~Quaglioni, and F.~Pederiva, ``{Optimal Control for the Quantum Simulation
  of Nuclear Dynamics},''
  \href{http://dx.doi.org/10.1103/PhysRevA.101.062307}{{\em Phys. Rev. A}
  {\bfseries 101} no.~6, (2020) 062307},
  \href{http://arxiv.org/abs/1908.08222}{{\ttfamily arXiv:1908.08222
  [quant-ph]}}.

\bibitem{Avkhadiev:2019niu}
A.~Avkhadiev, P.~Shanahan, and R.~Young, ``{Accelerating Lattice Quantum Field
  Theory Calculations via Interpolator Optimization Using Noisy
  Intermediate-Scale Quantum Computing},''
  \href{http://dx.doi.org/10.1103/PhysRevLett.124.080501}{{\em Phys. Rev.
  Lett.} {\bfseries 124} no.~8, (2020) 080501},
  \href{http://arxiv.org/abs/1908.04194}{{\ttfamily arXiv:1908.04194
  [hep-lat]}}.

\bibitem{Shaw:2020udc}
A.~F. Shaw, P.~Lougovski, J.~R. Stryker, and N.~Wiebe, ``{Quantum Algorithms
  for Simulating the Lattice Schwinger Model},''
  \href{http://dx.doi.org/10.22331/q-2020-08-10-306}{{\em Quantum} {\bfseries
  4} (2020) 306}, \href{http://arxiv.org/abs/2002.11146}{{\ttfamily
  arXiv:2002.11146 [quant-ph]}}.

\bibitem{Liu:2020eoa}
J.~Liu and Y.~Xin, ``{Quantum simulation of quantum field theories as quantum
  chemistry},'' \href{http://dx.doi.org/10.1007/JHEP12(2020)011}{{\em JHEP}
  {\bfseries 12} (2020) 011}, \href{http://arxiv.org/abs/2004.13234}{{\ttfamily
  arXiv:2004.13234 [hep-th]}}.

\bibitem{Kreshchuk:2020dla}
M.~Kreshchuk, W.~M. Kirby, G.~Goldstein, H.~Beauchemin, and P.~J. Love,
  ``{Quantum Simulation of Quantum Field Theory in the Light-Front
  Formulation},'' \href{http://arxiv.org/abs/2002.04016}{{\ttfamily
  arXiv:2002.04016 [quant-ph]}}.

\bibitem{Kharzeev:2020kgc}
D.~E. Kharzeev and Y.~Kikuchi, ``{Real-time chiral dynamics from a digital
  quantum simulation},''
  \href{http://dx.doi.org/10.1103/PhysRevResearch.2.023342}{{\em Phys. Rev.
  Res.} {\bfseries 2} no.~2, (2020) 023342},
  \href{http://arxiv.org/abs/2001.00698}{{\ttfamily arXiv:2001.00698
  [hep-ph]}}.

\bibitem{Klco:2020aud}
N.~Klco and M.~J. Savage, ``{Fixed-point quantum circuits for quantum field
  theories},'' \href{http://dx.doi.org/10.1103/PhysRevA.102.052422}{{\em Phys.
  Rev. A} {\bfseries 102} no.~5, (2020) 052422},
  \href{http://arxiv.org/abs/2002.02018}{{\ttfamily arXiv:2002.02018
  [quant-ph]}}.

\bibitem{DiMatteo:2020dhe}
O.~Di~Matteo, A.~Mccoy, P.~Gysbers, T.~Miyagi, R.~M. Woloshyn, and
  P.~Navr\'atil, ``{Improving Hamiltonian encodings with the Gray code},''
  \href{http://dx.doi.org/10.1103/PhysRevA.103.042405}{{\em Phys. Rev. A}
  {\bfseries 103} no.~4, (2021) 042405},
  \href{http://arxiv.org/abs/2008.05012}{{\ttfamily arXiv:2008.05012
  [quant-ph]}}.

\bibitem{Davoudi:2020yln}
Z.~Davoudi, I.~Raychowdhury, and A.~Shaw, ``{Search for Efficient Formulations
  for Hamiltonian Simulation of non-Abelian Lattice Gauge Theories},''
  \href{http://arxiv.org/abs/2009.11802}{{\ttfamily arXiv:2009.11802
  [hep-lat]}}.

\bibitem{Bepari:2020xqi}
K.~Bepari, S.~Malik, M.~Spannowsky, and S.~Williams, ``{Towards a quantum
  computing algorithm for helicity amplitudes and parton showers},''
  \href{http://dx.doi.org/10.1103/PhysRevD.103.076020}{{\em Phys. Rev. D}
  {\bfseries 103} no.~7, (2021) 076020},
  \href{http://arxiv.org/abs/2010.00046}{{\ttfamily arXiv:2010.00046
  [hep-ph]}}.

\bibitem{KOSSAKOWSKI1972247}
A.~Kossakowski, ``On quantum statistical mechanics of non-hamiltonian
  systems,''
  \href{http://dx.doi.org/https://doi.org/10.1016/0034-4877(72)90010-9}{{\em
  Reports on Mathematical Physics} {\bfseries 3} no.~4, (1972) 247 -- 274}.

\bibitem{Lindblad:1975ef}
G.~Lindblad, ``{On the Generators of Quantum Dynamical Semigroups},''
  \href{http://dx.doi.org/10.1007/BF01608499}{{\em Commun. Math. Phys.}
  {\bfseries 48} (1976) 119}.

\bibitem{Gorini:1976cm}
V.~Gorini, A.~Frigerio, M.~Verri, A.~Kossakowski, and E.~Sudarshan,
  ``{Properties of Quantum Markovian Master Equations},''
  \href{http://dx.doi.org/10.1016/0034-4877(78)90050-2}{{\em Rept. Math. Phys.}
  {\bfseries 13} (1978) 149}.

\bibitem{PhysRevA.83.062317}
H.~Wang, S.~Ashhab, and F.~Nori, ``Quantum algorithm for simulating the
  dynamics of an open quantum system,''
  \href{http://dx.doi.org/10.1103/PhysRevA.83.062317}{{\em Phys. Rev. A}
  {\bfseries 83} (Jun, 2011) 062317}.

\bibitem{PhysRevA.91.062308}
R.~Sweke, I.~Sinayskiy, D.~Bernard, and F.~Petruccione, ``Universal simulation
  of markovian open quantum systems,''
  \href{http://dx.doi.org/10.1103/PhysRevA.91.062308}{{\em Phys. Rev. A}
  {\bfseries 91} (Jun, 2015) 062308}.

\bibitem{Wei:2016}
S.-J. Wei, D.~Ruan, and G.-L. Long, ``Duality quantum algorithm efficiently
  simulates open quantum systems,''
  \href{http://dx.doi.org/10.1038/srep30727}{{\em Scientific Reports}
  {\bfseries 6} no.~1, (2016) 30727}.

\bibitem{cleve_et_al:LIPIcs:2017:7477}
R.~Cleve and C.~Wang, ``{Efficient Quantum Algorithms for Simulating Lindblad
  Evolution},'' in {\em 44th International Colloquium on Automata, Languages,
  and Programming (ICALP 2017)}.
\newblock \href{http://arxiv.org/abs/1612.09512}{{\ttfamily arXiv:1612.09512
  [quant-ph]}}.

\bibitem{PhysRevLett.118.140403}
A.~Chenu, M.~Beau, J.~Cao, and A.~del Campo, ``Quantum simulation of generic
  many-body open system dynamics using classical noise,''
  \href{http://dx.doi.org/10.1103/PhysRevLett.118.140403}{{\em Phys. Rev.
  Lett.} {\bfseries 118} (Apr, 2017) 140403}.

\bibitem{PhysRevA.101.012328}
H.-Y. Su and Y.~Li, ``Quantum algorithm for the simulation of open-system
  dynamics and thermalization,''
  \href{http://dx.doi.org/10.1103/PhysRevA.101.012328}{{\em Phys. Rev. A}
  {\bfseries 101} (Jan, 2020) 012328}.

\bibitem{PhysRevResearch.2.023214}
M.~Metcalf, J.~E. Moussa, W.~A. de~Jong, and M.~Sarovar, ``Engineered
  thermalization and cooling of quantum many-body systems,''
  \href{http://dx.doi.org/10.1103/PhysRevResearch.2.023214}{{\em Phys. Rev.
  Research} {\bfseries 2} (May, 2020) 023214}.
  \url{https://link.aps.org/doi/10.1103/PhysRevResearch.2.023214}.

\bibitem{Hu:2019}
Z.~Hu, R.~Xia, and S.~Kais, ``A quantum algorithm for evolving open quantum
  dynamics on quantum computing devices,''
  \href{http://dx.doi.org/10.1038/s41598-020-60321-x}{{\em Scientific Reports}
  {\bfseries 10} no.~1, (2020) 3301}.

\bibitem{PhysRevD.27.140}
J.~D. Bjorken, ``Highly relativistic nucleus-nucleus collisions: The central
  rapidity region,'' \href{http://dx.doi.org/10.1103/PhysRevD.27.140}{{\em
  Phys. Rev. D} {\bfseries 27} (Jan, 1983) 140--151}.

\bibitem{Arsene:2004fa}
{\bfseries BRAHMS} Collaboration, I.~Arsene {\em et~al.}, ``{Quark gluon plasma
  and color glass condensate at RHIC? The Perspective from the BRAHMS
  experiment},'' \href{http://dx.doi.org/10.1016/j.nuclphysa.2005.02.130}{{\em
  Nucl. Phys. A} {\bfseries 757} (2005) 1--27},
  \href{http://arxiv.org/abs/nucl-ex/0410020}{{\ttfamily
  arXiv:nucl-ex/0410020}}.

\bibitem{Adcox:2004mh}
{\bfseries PHENIX} Collaboration, K.~Adcox {\em et~al.}, ``{Formation of dense
  partonic matter in relativistic nucleus-nucleus collisions at RHIC:
  Experimental evaluation by the PHENIX collaboration},''
  \href{http://dx.doi.org/10.1016/j.nuclphysa.2005.03.086}{{\em Nucl. Phys. A}
  {\bfseries 757} (2005) 184--283},
  \href{http://arxiv.org/abs/nucl-ex/0410003}{{\ttfamily
  arXiv:nucl-ex/0410003}}.

\bibitem{Back:2004je}
{\bfseries PHOBOS} Collaboration, B.~Back {\em et~al.}, ``{The PHOBOS
  perspective on discoveries at RHIC},''
  \href{http://dx.doi.org/10.1016/j.nuclphysa.2005.03.084}{{\em Nucl. Phys. A}
  {\bfseries 757} (2005) 28--101},
  \href{http://arxiv.org/abs/nucl-ex/0410022}{{\ttfamily
  arXiv:nucl-ex/0410022}}.

\bibitem{Adams:2005dq}
{\bfseries STAR} Collaboration, J.~Adams {\em et~al.}, ``{Experimental and
  theoretical challenges in the search for the quark gluon plasma: The STAR
  Collaboration's critical assessment of the evidence from RHIC collisions},''
  \href{http://dx.doi.org/10.1016/j.nuclphysa.2005.03.085}{{\em Nucl. Phys. A}
  {\bfseries 757} (2005) 102--183},
  \href{http://arxiv.org/abs/nucl-ex/0501009}{{\ttfamily
  arXiv:nucl-ex/0501009}}.

\bibitem{LHC1review}
B.~M{\"u}ller, J.~Schukraft, and B.~Wys{\l}ouch, ``{First Results from Pb+Pb
  Collisions at the LHC},''
  \href{http://dx.doi.org/10.1146/annurev-nucl-102711-094910}{{\em Annu. Rev.
  Nucl. Part. S.} {\bfseries 62} no.~1, (2012) 361--386}.

\bibitem{Braun-Munzinger:2015hba}
P.~Braun-Munzinger, V.~Koch, T.~Sch{\"a}fer, and J.~Stachel, ``{Properties of
  hot and dense matter from relativistic heavy ion collisions},''
  \href{http://dx.doi.org/10.1016/j.physrep.2015.12.003}{{\em Phys. Rept.}
  {\bfseries 621} (2016) 76--126}.

\bibitem{TheBigPicture}
W.~Busza, K.~Rajagopal, and W.~van~der Schee, ``{Heavy Ion Collisions: The Big
  Picture, and the Big Questions},''
  \href{http://dx.doi.org/10.1146/annurev-nucl-101917-020852}{{\em Ann. Rev.
  Nucl. Part. Sci.} {\bfseries 68} (2018) 339--376}.

\bibitem{Weinberg:1977ji}
S.~Weinberg, {\em {The First Three Minutes. A Modern View of the Origin of the
  Universe}}.
\newblock ISBN:~9780465024377, Bantam Books, 1977.

\bibitem{Adare:2010de}
{\bfseries PHENIX} Collaboration, A.~Adare {\em et~al.}, ``{Heavy Quark
  Production in $p+p$ and Energy Loss and Flow of Heavy Quarks in Au+Au
  Collisions at $\sqrt{s_{NN}}=200$ GeV},''
  \href{http://dx.doi.org/10.1103/PhysRevC.84.044905}{{\em Phys. Rev. C}
  {\bfseries 84} (2011) 044905},
  \href{http://arxiv.org/abs/1005.1627}{{\ttfamily arXiv:1005.1627 [nucl-ex]}}.

\bibitem{Sirunyan:2017isk}
{\bfseries CMS} Collaboration, A.~M. Sirunyan {\em et~al.}, ``{Measurement of
  prompt and nonprompt charmonium suppression in $\text {PbPb}$ collisions at
  5.02 $\,\text {Te}\text {V}$},''
  \href{http://dx.doi.org/10.1140/epjc/s10052-018-5950-6}{{\em Eur. Phys. J. C}
  {\bfseries 78} no.~6, (2018) 509},
  \href{http://arxiv.org/abs/1712.08959}{{\ttfamily arXiv:1712.08959
  [nucl-ex]}}.

\bibitem{Adamczyk:2017yhe}
{\bfseries STAR} Collaboration, L.~Adamczyk {\em et~al.}, ``{Measurements of
  jet quenching with semi-inclusive hadron+jet distributions in Au+Au
  collisions at $\sqrt{s_{NN}}$ = 200 GeV},''
  \href{http://dx.doi.org/10.1103/PhysRevC.96.024905}{{\em Phys. Rev. C}
  {\bfseries 96} no.~2, (2017) 024905},
  \href{http://arxiv.org/abs/1702.01108}{{\ttfamily arXiv:1702.01108
  [nucl-ex]}}.

\bibitem{Acharya:2019jyg}
{\bfseries ALICE} Collaboration, S.~Acharya {\em et~al.}, ``{Measurements of
  inclusive jet spectra in pp and central Pb-Pb collisions at
  $\sqrt{s_{\rm{NN}}}$ = 5.02 TeV},''
  \href{http://dx.doi.org/10.1103/PhysRevC.101.034911}{{\em Phys. Rev. C}
  {\bfseries 101} no.~3, (2020) 034911},
  \href{http://arxiv.org/abs/1909.09718}{{\ttfamily arXiv:1909.09718
  [nucl-ex]}}.

\bibitem{Aaboud:2018twu}
{\bfseries ATLAS} Collaboration, M.~Aaboud {\em et~al.}, ``{Measurement of the
  nuclear modification factor for inclusive jets in Pb+Pb collisions at
  $\sqrt{s_\mathrm{NN}}=5.02$ TeV with the ATLAS detector},''
  \href{http://dx.doi.org/10.1016/j.physletb.2018.10.076}{{\em Phys. Lett. B}
  {\bfseries 790} (2019) 108--128},
  \href{http://arxiv.org/abs/1805.05635}{{\ttfamily arXiv:1805.05635
  [nucl-ex]}}.

\bibitem{Gossiaux:2008jv}
P.~Gossiaux and J.~Aichelin, ``{Towards an understanding of the RHIC single
  electron data},'' \href{http://dx.doi.org/10.1103/PhysRevC.78.014904}{{\em
  Phys. Rev. C} {\bfseries 78} (2008) 014904},
  \href{http://arxiv.org/abs/0802.2525}{{\ttfamily arXiv:0802.2525 [hep-ph]}}.

\bibitem{Schenke:2009gb}
B.~Schenke, C.~Gale, and S.~Jeon, ``{MARTINI: An Event generator for
  relativistic heavy-ion collisions},''
  \href{http://dx.doi.org/10.1103/PhysRevC.80.054913}{{\em Phys. Rev. C}
  {\bfseries 80} (2009) 054913},
  \href{http://arxiv.org/abs/0909.2037}{{\ttfamily arXiv:0909.2037 [hep-ph]}}.

\bibitem{Wang:2013cia}
X.-N. Wang and Y.~Zhu, ``{Medium Modification of $\gamma$-jets in High-energy
  Heavy-ion Collisions},''
  \href{http://dx.doi.org/10.1103/PhysRevLett.111.062301}{{\em Phys. Rev.
  Lett.} {\bfseries 111} no.~6, (2013) 062301},
  \href{http://arxiv.org/abs/1302.5874}{{\ttfamily arXiv:1302.5874 [hep-ph]}}.

\bibitem{Cao:2016gvr}
S.~Cao, T.~Luo, G.-Y. Qin, and X.-N. Wang, ``{Linearized Boltzmann transport
  model for jet propagation in the quark-gluon plasma: Heavy quark
  evolution},'' \href{http://dx.doi.org/10.1103/PhysRevC.94.014909}{{\em Phys.
  Rev. C} {\bfseries 94} no.~1, (2016) 014909},
  \href{http://arxiv.org/abs/1605.06447}{{\ttfamily arXiv:1605.06447
  [nucl-th]}}.

\bibitem{Cao:2015hia}
S.~Cao, G.-Y. Qin, and S.~A. Bass, ``{Energy loss, hadronization and hadronic
  interactions of heavy flavors in relativistic heavy-ion collisions},''
  \href{http://dx.doi.org/10.1103/PhysRevC.92.024907}{{\em Phys. Rev. C}
  {\bfseries 92} no.~2, (2015) 024907},
  \href{http://arxiv.org/abs/1505.01413}{{\ttfamily arXiv:1505.01413
  [nucl-th]}}.

\bibitem{Du:2017qkv}
X.~Du, R.~Rapp, and M.~He, ``{Color Screening and Regeneration of Bottomonia in
  High-Energy Heavy-Ion Collisions},''
  \href{http://dx.doi.org/10.1103/PhysRevC.96.054901}{{\em Phys. Rev. C}
  {\bfseries 96} no.~5, (2017) 054901},
  \href{http://arxiv.org/abs/1706.08670}{{\ttfamily arXiv:1706.08670
  [hep-ph]}}.

\bibitem{Ke:2018tsh}
W.~Ke, Y.~Xu, and S.~A. Bass, ``{Linearized Boltzmann-Langevin model for heavy
  quark transport in hot and dense QCD matter},''
  \href{http://dx.doi.org/10.1103/PhysRevC.98.064901}{{\em Phys. Rev. C}
  {\bfseries 98} no.~6, (2018) 064901},
  \href{http://arxiv.org/abs/1806.08848}{{\ttfamily arXiv:1806.08848
  [nucl-th]}}.

\bibitem{Yao:2020xzw}
X.~Yao, W.~Ke, Y.~Xu, S.~A. Bass, and B.~M\"uller, ``{Coupled Boltzmann
  Transport Equations of Heavy Quarks and Quarkonia in Quark-Gluon Plasma},''
  \href{http://dx.doi.org/10.1007/JHEP01(2021)046}{{\em JHEP} {\bfseries 01}
  (2021) 046}, \href{http://arxiv.org/abs/2004.06746}{{\ttfamily
  arXiv:2004.06746 [hep-ph]}}.

\bibitem{Blaizot:2017ypk}
J.-P. Blaizot and M.~A. Escobedo, ``{Quantum and classical dynamics of heavy
  quarks in a quark-gluon plasma},''
  \href{http://dx.doi.org/10.1007/JHEP06(2018)034}{{\em JHEP} {\bfseries 06}
  (2018) 034}, \href{http://arxiv.org/abs/1711.10812}{{\ttfamily
  arXiv:1711.10812 [hep-ph]}}.

\bibitem{Yao:2020eqy}
X.~Yao and T.~Mehen, ``{Quarkonium Semiclassical Transport in Quark-Gluon
  Plasma: Factorization and Quantum Correction},''
  \href{http://dx.doi.org/10.1007/JHEP02(2021)062}{{\em JHEP} {\bfseries 02}
  (2021) 062}, \href{http://arxiv.org/abs/2009.02408}{{\ttfamily
  arXiv:2009.02408 [hep-ph]}}.

\bibitem{Young:2010jq}
C.~Young and K.~Dusling, ``{Quarkonium above deconfinement as an open quantum
  system},'' \href{http://dx.doi.org/10.1103/PhysRevC.87.065206}{{\em Phys.
  Rev. C} {\bfseries 87} no.~6, (2013) 065206},
  \href{http://arxiv.org/abs/1001.0935}{{\ttfamily arXiv:1001.0935 [nucl-th]}}.

\bibitem{Akamatsu:2011se}
Y.~Akamatsu and A.~Rothkopf, ``{Stochastic potential and quantum decoherence of
  heavy quarkonium in the quark-gluon plasma},''
  \href{http://dx.doi.org/10.1103/PhysRevD.85.105011}{{\em Phys. Rev. D}
  {\bfseries 85} (2012) 105011},
  \href{http://arxiv.org/abs/1110.1203}{{\ttfamily arXiv:1110.1203 [hep-ph]}}.

\bibitem{Gossiaux:2016htk}
P.~B. Gossiaux and R.~Katz, ``{Upsilon suppression in the
  Schrödinger--Langevin approach},''
  \href{http://dx.doi.org/10.1016/j.nuclphysa.2016.04.017}{{\em Nucl. Phys. A}
  {\bfseries 956} (2016) 737--740},
  \href{http://arxiv.org/abs/1601.01443}{{\ttfamily arXiv:1601.01443
  [hep-ph]}}.

\bibitem{Brambilla:2017zei}
N.~Brambilla, M.~A. Escobedo, J.~Soto, and A.~Vairo, ``{Heavy quarkonium
  suppression in a fireball},''
  \href{http://dx.doi.org/10.1103/PhysRevD.97.074009}{{\em Phys. Rev. D}
  {\bfseries 97} no.~7, (2018) 074009},
  \href{http://arxiv.org/abs/1711.04515}{{\ttfamily arXiv:1711.04515
  [hep-ph]}}.

\bibitem{Yao:2018nmy}
X.~Yao and T.~Mehen, ``{Quarkonium in-medium transport equation derived from
  first principles},'' \href{http://dx.doi.org/10.1103/PhysRevD.99.096028}{{\em
  Phys. Rev. D} {\bfseries 99} no.~9, (2019) 096028},
  \href{http://arxiv.org/abs/1811.07027}{{\ttfamily arXiv:1811.07027
  [hep-ph]}}.

\bibitem{Miura:2019ssi}
T.~Miura, Y.~Akamatsu, M.~Asakawa, and A.~Rothkopf, ``{Quantum Brownian motion
  of a heavy quark pair in the quark-gluon plasma},''
  \href{http://dx.doi.org/10.1103/PhysRevD.101.034011}{{\em Phys. Rev. D}
  {\bfseries 101} no.~3, (2020) 034011},
  \href{http://arxiv.org/abs/1908.06293}{{\ttfamily arXiv:1908.06293
  [nucl-th]}}.

\bibitem{Sharma:2019xum}
R.~Sharma and A.~Tiwari, ``{Quantum evolution of quarkonia with correlated and
  uncorrelated noise},''
  \href{http://dx.doi.org/10.1103/PhysRevD.101.074004}{{\em Phys. Rev. D}
  {\bfseries 101} no.~7, (2020) 074004},
  \href{http://arxiv.org/abs/1912.07036}{{\ttfamily arXiv:1912.07036
  [hep-ph]}}.

\bibitem{Brambilla:2020qwo}
N.~Brambilla, M.~A. Escobedo, M.~Strickland, A.~Vairo, P.~Vander~Griend, and
  J.~H. Weber, ``{Bottomonium suppression in an open quantum system using the
  quantum trajectories method},''
  \href{http://dx.doi.org/10.1007/JHEP05(2021)136}{{\em JHEP} {\bfseries 05}
  (2021) 136}, \href{http://arxiv.org/abs/2012.01240}{{\ttfamily
  arXiv:2012.01240 [hep-ph]}}.

\bibitem{Vaidya:2020cyi}
V.~Vaidya and X.~Yao, ``{Transverse momentum broadening of a jet in quark-gluon
  plasma: an open quantum system EFT},''
  \href{http://dx.doi.org/10.1007/JHEP10(2020)024}{{\em JHEP} {\bfseries 10}
  (2020) 024}, \href{http://arxiv.org/abs/2004.11403}{{\ttfamily
  arXiv:2004.11403 [hep-ph]}}.

\bibitem{Vaidya:2020lih}
V.~Vaidya, ``{Effective Field Theory for jet substructure in heavy ion
  collisions},'' \href{http://arxiv.org/abs/2010.00028}{{\ttfamily
  arXiv:2010.00028 [hep-ph]}}.

\bibitem{Andronic:2007bi}
A.~Andronic, P.~Braun-Munzinger, K.~Redlich, and J.~Stachel, ``{Evidence for
  charmonium generation at the phase boundary in ultra-relativistic nuclear
  collisions},'' \href{http://dx.doi.org/10.1016/j.physletb.2007.07.036}{{\em
  Phys. Lett. B} {\bfseries 652} (2007) 259--261},
  \href{http://arxiv.org/abs/nucl-th/0701079}{{\ttfamily
  arXiv:nucl-th/0701079}}.

\bibitem{Abada:2019lih}
{\bfseries FCC} Collaboration, A.~Abada {\em et~al.}, ``{FCC Physics
  Opportunities}: {Future Circular Collider Conceptual Design Report Volume
  1},'' \href{http://dx.doi.org/10.1140/epjc/s10052-019-6904-3}{{\em Eur. Phys.
  J. C} {\bfseries 79} no.~6, (2019) 474}.

\bibitem{Jordan:2011ci}
S.~P. Jordan, K.~S.~M. Lee, and J.~Preskill, ``{Quantum Computation of
  Scattering in Scalar Quantum Field Theories},'' {\em Quant. Inf. Comput.}
  {\bfseries 14} (2014) 1014--1080,
  \href{http://arxiv.org/abs/1112.4833}{{\ttfamily arXiv:1112.4833 [hep-th]}}.

\bibitem{Martinez_2016}
E.~A. Martinez, C.~A. Muschik, P.~Schindler, D.~Nigg, A.~Erhard, M.~Heyl,
  P.~Hauke, M.~Dalmonte, T.~Monz, P.~Zoller, and et~al., ``Real-time dynamics
  of lattice gauge theories with a few-qubit quantum computer,''
  \href{http://dx.doi.org/10.1038/nature18318}{{\em Nature} {\bfseries 534}
  no.~7608, (Jun, 2016) 516–519}.

\bibitem{Haase:2020kaj}
J.~F. Haase, L.~Dellantonio, A.~Celi, D.~Paulson, A.~Kan, K.~Jansen, and C.~A.
  Muschik, ``{A resource efficient approach for quantum and classical
  simulations of gauge theories in particle physics},''
  \href{http://dx.doi.org/10.22331/q-2021-02-04-393}{{\em Quantum} {\bfseries
  5} (2021) 393}, \href{http://arxiv.org/abs/2006.14160}{{\ttfamily
  arXiv:2006.14160 [quant-ph]}}.

\bibitem{CasalderreySolana:2004qm}
J.~Casalderrey-Solana, E.~Shuryak, and D.~Teaney, ``{Conical flow induced by
  quenched QCD jets},''
  \href{http://dx.doi.org/10.1088/1742-6596/27/1/003}{{\em J. Phys. Conf. Ser.}
  {\bfseries 27} (2005) 22--31},
  \href{http://arxiv.org/abs/hep-ph/0411315}{{\ttfamily arXiv:hep-ph/0411315}}.

\bibitem{Ruppert:2005uz}
J.~Ruppert and B.~Müller, ``{Waking the colored plasma},''
  \href{http://dx.doi.org/10.1016/j.physletb.2005.04.075}{{\em Phys. Lett. B}
  {\bfseries 618} (2005) 123--130},
  \href{http://arxiv.org/abs/hep-ph/0503158}{{\ttfamily arXiv:hep-ph/0503158}}.

\bibitem{Chaudhuri:2005vc}
A.~Chaudhuri and U.~Heinz, ``{Effect of jet quenching on the hydrodynamical
  evolution of QGP},''
  \href{http://dx.doi.org/10.1103/PhysRevLett.97.062301}{{\em Phys. Rev. Lett.}
  {\bfseries 97} (2006) 062301},
  \href{http://arxiv.org/abs/nucl-th/0503028}{{\ttfamily
  arXiv:nucl-th/0503028}}.

\bibitem{Qin:2009uh}
G.-Y. Qin, A.~Majumder, H.~Song, and U.~Heinz, ``{Energy and momentum deposited
  into a QCD medium by a jet shower},''
  \href{http://dx.doi.org/10.1103/PhysRevLett.103.152303}{{\em Phys. Rev.
  Lett.} {\bfseries 103} (2009) 152303},
  \href{http://arxiv.org/abs/0903.2255}{{\ttfamily arXiv:0903.2255 [nucl-th]}}.

\bibitem{Betz:2010qh}
B.~Betz, J.~Noronha, G.~Torrieri, M.~Gyulassy, and D.~H. Rischke, ``{Universal
  Flow-Driven Conical Emission in Ultrarelativistic Heavy-Ion Collisions},''
  \href{http://dx.doi.org/10.1103/PhysRevLett.105.222301}{{\em Phys. Rev.
  Lett.} {\bfseries 105} (2010) 222301},
  \href{http://arxiv.org/abs/1005.5461}{{\ttfamily arXiv:1005.5461 [nucl-th]}}.

\bibitem{Floerchinger:2014yqa}
S.~Floerchinger and K.~C. Zapp, ``{Hydrodynamics and Jets in Dialogue},''
  \href{http://dx.doi.org/10.1140/epjc/s10052-014-3189-4}{{\em Eur. Phys. J. C}
  {\bfseries 74} no.~12, (2014) 3189},
  \href{http://arxiv.org/abs/1407.1782}{{\ttfamily arXiv:1407.1782 [hep-ph]}}.

\bibitem{Chen:2017zte}
W.~Chen, S.~Cao, T.~Luo, L.-G. Pang, and X.-N. Wang, ``{Effects of jet-induced
  medium excitation in $\gamma$-hadron correlation in A+A collisions},''
  \href{http://dx.doi.org/10.1016/j.physletb.2017.12.015}{{\em Phys. Lett. B}
  {\bfseries 777} (2018) 86--90},
  \href{http://arxiv.org/abs/1704.03648}{{\ttfamily arXiv:1704.03648
  [nucl-th]}}.

\bibitem{Yan:2017rku}
L.~Yan, S.~Jeon, and C.~Gale, ``{Jet-medium interaction and conformal
  relativistic fluid dynamics},''
  \href{http://dx.doi.org/10.1103/PhysRevC.97.034914}{{\em Phys. Rev. C}
  {\bfseries 97} no.~3, (2018) 034914},
  \href{http://arxiv.org/abs/1707.09519}{{\ttfamily arXiv:1707.09519
  [nucl-th]}}.

\bibitem{Tachibana:2020mtb}
Y.~Tachibana, C.~Shen, and A.~Majumder, ``{Bulk medium evolution has
  considerable effects on jet observables!},''
  \href{http://arxiv.org/abs/2001.08321}{{\ttfamily arXiv:2001.08321
  [nucl-th]}}.

\bibitem{Casalderrey-Solana:2020rsj}
J.~Casalderrey-Solana, J.~G. Milhano, D.~Pablos, K.~Rajagopal, and X.~Yao,
  ``{Jet Wake from Linearized Hydrodynamics},''
  \href{http://dx.doi.org/10.1007/JHEP05(2021)230}{{\em JHEP} {\bfseries 05}
  (2021) 230}, \href{http://arxiv.org/abs/2010.01140}{{\ttfamily
  arXiv:2010.01140 [hep-ph]}}.

\bibitem{Cao:2020wlm}
S.~Cao and X.-N. Wang, ``{Jet quenching and medium response in high-energy
  heavy-ion collisions: a review},''
  \href{http://dx.doi.org/10.1088/1361-6633/abc22b}{{\em Rept. Prog. Phys.}
  {\bfseries 84} no.~2, (2021) 024301},
  \href{http://arxiv.org/abs/2002.04028}{{\ttfamily arXiv:2002.04028
  [hep-ph]}}.

\bibitem{Petreczky:2005nh}
P.~Petreczky and D.~Teaney, ``{Heavy quark diffusion from the lattice},''
  \href{http://dx.doi.org/10.1103/PhysRevD.73.014508}{{\em Phys. Rev. D}
  {\bfseries 73} (2006) 014508},
  \href{http://arxiv.org/abs/hep-ph/0507318}{{\ttfamily arXiv:hep-ph/0507318}}.

\bibitem{Banerjee:2011ra}
D.~Banerjee, S.~Datta, R.~Gavai, and P.~Majumdar, ``{Heavy Quark Momentum
  Diffusion Coefficient from Lattice QCD},''
  \href{http://dx.doi.org/10.1103/PhysRevD.85.014510}{{\em Phys. Rev. D}
  {\bfseries 85} (2012) 014510},
  \href{http://arxiv.org/abs/1109.5738}{{\ttfamily arXiv:1109.5738 [hep-lat]}}.

\bibitem{Majumder:2012sh}
A.~Majumder, ``{Calculating the jet quenching parameter $\hat{q}$ in lattice
  gauge theory},'' \href{http://dx.doi.org/10.1103/PhysRevC.87.034905}{{\em
  Phys. Rev. C} {\bfseries 87} (2013) 034905},
  \href{http://arxiv.org/abs/1202.5295}{{\ttfamily arXiv:1202.5295 [nucl-th]}}.

\bibitem{Francis:2015daa}
A.~Francis, O.~Kaczmarek, M.~Laine, T.~Neuhaus, and H.~Ohno, ``{Nonperturbative
  estimate of the heavy quark momentum diffusion coefficient},''
  \href{http://dx.doi.org/10.1103/PhysRevD.92.116003}{{\em Phys. Rev. D}
  {\bfseries 92} no.~11, (2015) 116003},
  \href{http://arxiv.org/abs/1508.04543}{{\ttfamily arXiv:1508.04543
  [hep-lat]}}.

\bibitem{Brambilla:2020siz}
N.~Brambilla, V.~Leino, P.~Petreczky, and A.~Vairo, ``{Lattice QCD constraints
  on the heavy quark diffusion coefficient},''
  \href{http://dx.doi.org/10.1103/PhysRevD.102.074503}{{\em Phys. Rev. D}
  {\bfseries 102} no.~7, (2020) 074503},
  \href{http://arxiv.org/abs/2007.10078}{{\ttfamily arXiv:2007.10078
  [hep-lat]}}.

\bibitem{Liu:2006ug}
H.~Liu, K.~Rajagopal, and U.~A. Wiedemann, ``{Calculating the jet quenching
  parameter from AdS/CFT},''
  \href{http://dx.doi.org/10.1103/PhysRevLett.97.182301}{{\em Phys. Rev. Lett.}
  {\bfseries 97} (2006) 182301},
  \href{http://arxiv.org/abs/hep-ph/0605178}{{\ttfamily arXiv:hep-ph/0605178}}.

\bibitem{Liu:2006he}
H.~Liu, K.~Rajagopal, and U.~A. Wiedemann, ``{Wilson loops in heavy ion
  collisions and their calculation in AdS/CFT},''
  \href{http://dx.doi.org/10.1088/1126-6708/2007/03/066}{{\em JHEP} {\bfseries
  03} (2007) 066}, \href{http://arxiv.org/abs/hep-ph/0612168}{{\ttfamily
  arXiv:hep-ph/0612168}}.

\bibitem{CasalderreySolana:2006rq}
J.~Casalderrey-Solana and D.~Teaney, ``{Heavy quark diffusion in strongly
  coupled N=4 Yang-Mills},''
  \href{http://dx.doi.org/10.1103/PhysRevD.74.085012}{{\em Phys. Rev. D}
  {\bfseries 74} (2006) 085012},
  \href{http://arxiv.org/abs/hep-ph/0605199}{{\ttfamily arXiv:hep-ph/0605199}}.

\bibitem{CaronHuot:2008uh}
S.~Caron-Huot and G.~D. Moore, ``{Heavy quark diffusion in QCD and N=4 SYM at
  next-to-leading order},''
  \href{http://dx.doi.org/10.1088/1126-6708/2008/02/081}{{\em JHEP} {\bfseries
  02} (2008) 081}, \href{http://arxiv.org/abs/0801.2173}{{\ttfamily
  arXiv:0801.2173 [hep-ph]}}.

\bibitem{CasalderreySolana:2011us}
J.~Casalderrey-Solana, H.~Liu, D.~Mateos, K.~Rajagopal, and U.~A. Wiedemann,
  \href{http://dx.doi.org/10.1017/CBO9781139136747}{{\em {Gauge/String Duality,
  Hot QCD and Heavy Ion Collisions}}}.
\newblock Cambridge University Press, 2014.
\newblock \href{http://arxiv.org/abs/1101.0618}{{\ttfamily arXiv:1101.0618
  [hep-th]}}.

\bibitem{Brambilla:2019tpt}
N.~Brambilla, M.~A. Escobedo, A.~Vairo, and P.~Vander~Griend, ``{Transport
  coefficients from in medium quarkonium dynamics},''
  \href{http://dx.doi.org/10.1103/PhysRevD.100.054025}{{\em Phys. Rev. D}
  {\bfseries 100} no.~5, (2019) 054025},
  \href{http://arxiv.org/abs/1903.08063}{{\ttfamily arXiv:1903.08063
  [hep-ph]}}.

\bibitem{nielsen_chuang_2010}
M.~A. Nielsen and I.~L. Chuang,
  \href{http://dx.doi.org/10.1017/CBO9780511976667}{{\em Quantum Computation
  and Quantum Information: 10th Anniversary Edition}}.
\newblock Cambridge University Press, 2010.

\bibitem{Qiskit}
A.~e.~a. H{\'e}ctor~Abraham, ``Qiskit: An open-source framework for quantum
  computing,'' 2019.

\bibitem{younis2020qfast}
E.~Younis, K.~Sen, K.~Yelick, and C.~Iancu, ``Qfast: Quantum synthesis using a
  hierarchical continuous circuit space,''
  \href{http://arxiv.org/abs/2003.04462}{{\ttfamily arXiv:2003.04462
  [quant-ph]}}.

\bibitem{davis2019heuristics}
M.~G. Davis, E.~Smith, A.~Tudor, K.~Sen, I.~Siddiqi, and C.~Iancu, ``Heuristics
  for quantum compiling with a continuous gate set,''
  \href{http://arxiv.org/abs/1912.02727}{{\ttfamily arXiv:1912.02727 [cs.ET]}}.

\bibitem{qsearch_placeholder}
T.~E. Davis, A.~Tudor, K.~Sen, I.~Siddiqi, and C.~Iancu, ``Towards depth
  optimal topology aware quantum circuit synthesis,'' in {\em {IEEE
  International Conference on Quantum Computing and Engineering (QCE20)}}.
\newblock 2020.

\bibitem{IBMQVigo}
{5-qubit backend: IBM Q team, IBM Q 5 Vigo backend specification v1.2.1, (2020)
  Retrieved from
  \href{https://quantum-computing.ibm.com}{https://quantum-computing.ibm.com}}.

\bibitem{He:2020udd}
A.~He, B.~Nachman, W.~A. de~Jong, and C.~W. Bauer, ``{Zero-noise extrapolation
  for quantum-gate error mitigation with identity insertions},''
  \href{http://dx.doi.org/10.1103/PhysRevA.102.012426}{{\em Phys. Rev. A}
  {\bfseries 102} no.~1, (2020) 012426},
  \href{http://arxiv.org/abs/2003.04941}{{\ttfamily arXiv:2003.04941
  [quant-ph]}}.

\bibitem{IBMQValencia}
{5-qubit backend: IBM Q team, IBM Q 5 Valencia backend specification v1.3.1,
  (2020) Retrieved from
  \href{https://quantum-computing.ibm.com}{https://quantum-computing.ibm.com}}.

\bibitem{IBMQSantiago}
{5-qubit backend: IBM Q team, IBM Q 5 Santiago backend specification v1.0.3,
  (2020) Retrieved from
  \href{https://quantum-computing.ibm.com}{https://quantum-computing.ibm.com}}.

\bibitem{Accardi:2012qut}
A.~Accardi {\em et~al.}, ``{Electron Ion Collider: The Next QCD Frontier}:
  {Understanding the glue that binds us all},''
  \href{http://dx.doi.org/10.1140/epja/i2016-16268-9}{{\em Eur. Phys. J. A}
  {\bfseries 52} no.~9, (2016) 268},
  \href{http://arxiv.org/abs/1212.1701}{{\ttfamily arXiv:1212.1701 [nucl-ex]}}.

\bibitem{Dasgupta:2001sh}
M.~Dasgupta and G.~Salam, ``{Resummation of nonglobal QCD observables},''
  \href{http://dx.doi.org/10.1016/S0370-2693(01)00725-0}{{\em Phys. Lett. B}
  {\bfseries 512} (2001) 323--330},
  \href{http://arxiv.org/abs/hep-ph/0104277}{{\ttfamily arXiv:hep-ph/0104277}}.

\bibitem{Banfi:2002hw}
A.~Banfi, G.~Marchesini, and G.~Smye, ``{Away from jet energy flow},''
  \href{http://dx.doi.org/10.1088/1126-6708/2002/08/006}{{\em JHEP} {\bfseries
  08} (2002) 006}, \href{http://arxiv.org/abs/hep-ph/0206076}{{\ttfamily
  arXiv:hep-ph/0206076}}.

\bibitem{Nagy:2007ty}
Z.~Nagy and D.~E. Soper, ``{Parton showers with quantum interference},''
  \href{http://dx.doi.org/10.1088/1126-6708/2007/09/114}{{\em JHEP} {\bfseries
  09} (2007) 114}, \href{http://arxiv.org/abs/0706.0017}{{\ttfamily
  arXiv:0706.0017 [hep-ph]}}.

\bibitem{Neill:2015nya}
D.~Neill, ``{The Edge of Jets and Subleading Non-Global Logs},''
  \href{http://arxiv.org/abs/1508.07568}{{\ttfamily arXiv:1508.07568
  [hep-ph]}}.

\bibitem{Armesto:2019mna}
N.~Armesto, F.~Dominguez, A.~Kovner, M.~Lublinsky, and V.~Skokov, ``{The Color
  Glass Condensate density matrix: Lindblad evolution, entanglement entropy and
  Wigner functional},'' \href{http://dx.doi.org/10.1007/JHEP05(2019)025}{{\em
  JHEP} {\bfseries 05} (2019) 025},
  \href{http://arxiv.org/abs/1901.08080}{{\ttfamily arXiv:1901.08080
  [hep-ph]}}.

\bibitem{Li:2020bys}
M.~Li and A.~Kovner, ``{JIMWLK Evolution, Lindblad Equation and
  Quantum-Classical Correspondence},''
  \href{http://dx.doi.org/10.1007/JHEP05(2020)036}{{\em JHEP} {\bfseries 05}
  (2020) 036}, \href{http://arxiv.org/abs/2002.02282}{{\ttfamily
  arXiv:2002.02282 [hep-ph]}}.

\end{thebibliography}\endgroup

\widetext
\appendix

\vspace*{.5cm}
\section{Supplemental material}

\begin{figure}[h!]
  \vspace*{.3cm}
  \centerline{\includegraphics[width = 1 \textwidth]{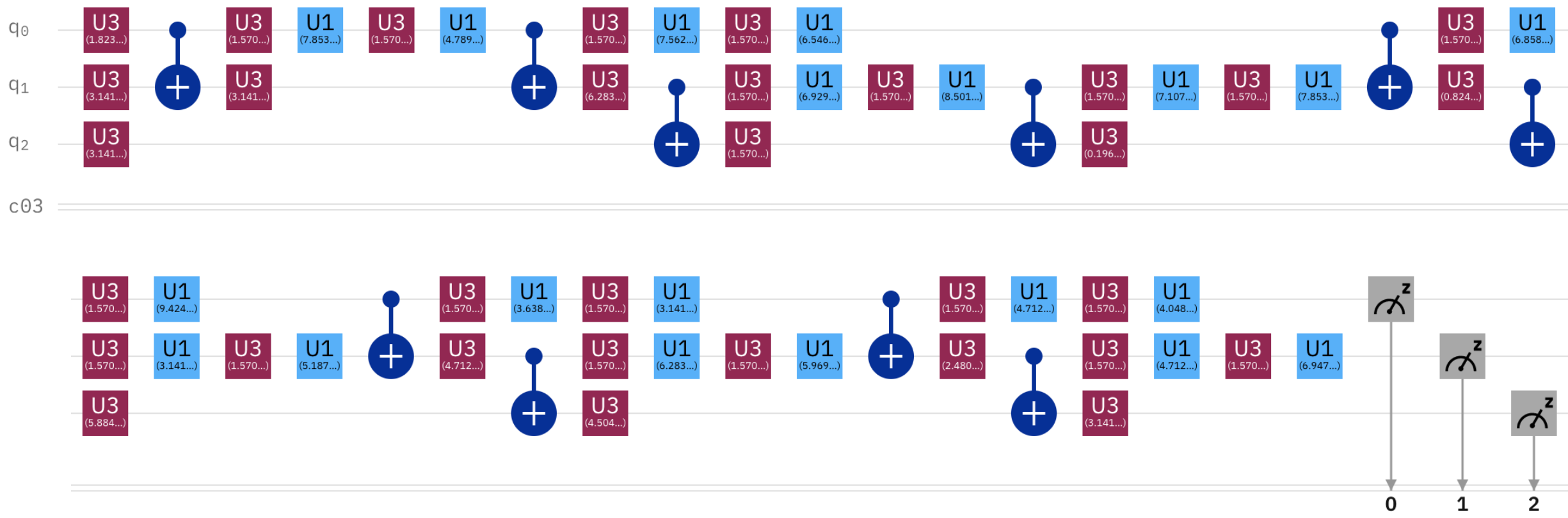}}
  \vspace*{.5cm}
  \caption{Decomposition of a single cycle of the quantum algorithm in Fig.~\ref{fig:circuit} in terms of single qubit rotations (U1,3) and \textsc{cnot} gates using the \texttt{qsearch} compiler of Ref.~\cite{davis2019heuristics}. Here \texttt{q\textsubscript{0}} corresponds to the system qubit, \texttt{q\textsubscript{1,2}} are the auxiliary qubits and \texttt{c03} represents three classical bits for the readout. The result of $P_0(t)$ from the final trace-out and measurement can be written as $P_0(t) = \sum_{i,j=0}^1 \langle 0 ij | \rho(t) | 0ij\rangle$ where $\langle 0 ij | \rho(t) | 0ij\rangle$ is the measurement result for $\texttt{q\textsubscript{0}}=0$, $\texttt{q\textsubscript{1}}=i$, $\texttt{q\textsubscript{2}}=j$. }
  \label{fig:circuit_qsearch}
\end{figure}

\begin{figure}[h!]
  \centerline{\includegraphics[width = 0.5 \textwidth]{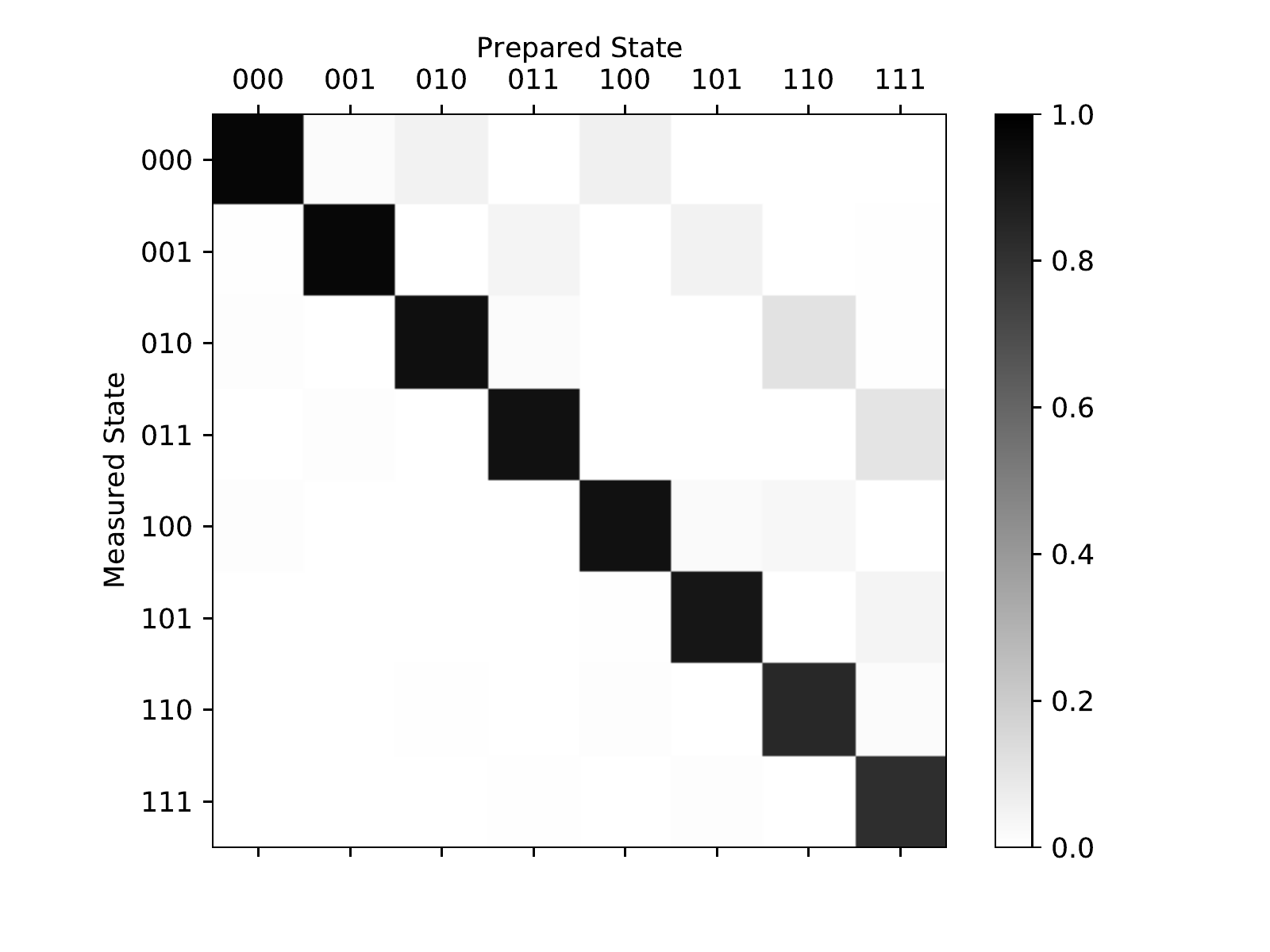}}
  \caption{The response matrix of the qubits \texttt{q\textsubscript{0-2}} of IBM~Q Vigo device~\cite{IBMQVigo} which is used for the readout error mitigation in Fig.~\ref{fig:device_result}. The $2^3$ states are prepared by applying $X$ gates and then corresponding measurements are performed. The error mitigation is implemented using the constrained matrix inversion approach which is implemented in IBM's \texttt{qiskit-ignis} package~\cite{Qiskit}.}
  \label{fig:circuit_qsearch}
\end{figure}

\end{document}